\documentclass[sn-mathphys,Numbered]{sn-jnl}

\usepackage{graphicx}%
\usepackage{multirow}%
\usepackage{amsmath,amssymb,amsfonts}%
\usepackage{amsthm}%
\usepackage{mathrsfs}%
\usepackage[title]{appendix}%
\usepackage{xcolor}%
\usepackage{textcomp}%
\usepackage{manyfoot}%
\usepackage{booktabs}%
\usepackage{algorithm}%
\usepackage{algorithmicx}%
\usepackage{algpseudocode}%
\usepackage{listings}%
\usepackage{titletoc}%
\usepackage{soul}%
\usepackage{textcomp}%
\usepackage{textgreek}
\usepackage{pifont}
\usepackage{geometry}
\usepackage[resetlabels]{multibib}

\newcites{S}{Supplementary Note References}

\usepackage{lineno}
\usepackage{acronym}
\usepackage{newfloat}
\DeclareFloatingEnvironment[fileext=lop]{Extended Data Fig.}
\DeclareFloatingEnvironment[fileext=lop]{Extended Data Table}

\acrodef{EO}{electro-optic}
\acrodef{HPC}{high-performance computer}
\acrodef{XR}{extended reality}
\acrodef{MPW}{multi-project wafer}
\acrodef{CMOS}{complementary metal-oxide-semiconductor}
\acrodef{CPO}{co-packaged optics}
\acrodef{AI}{artificial intelligence}
\acrodef{ML}{machine learning}
\acrodef{SiP}{silicon photonics}
\acrodef{SiN}{silicon nitride}
\acrodef{MZM}{Mach-Zehnder modulator}
\acrodef{MZI}{Mach-Zehnder interferometer}
\acrodef{MRM}{microring modulator}
\acrodef{MRA-MZM}{microring-assisted MZM}
\acrodef{MOS}{metal-oxide-semiconductor}
\acrodef{DUT}{device under test}
\acrodef{TOPS}{thermo-optic phase shifter}
\acrodef{TEC}{thermo-electric cooler}
\acrodef{FWHM}{full width at half-maximum}

\acrodef{DAC}{digital-to-analog converter}
\acrodef{DCI}{data-center interconnect}
\acrodef{ADC}{analog-to-digital converter}
\acrodef{AWG}{arbitrary waveform generator}
\acrodef{I/Q}{in-phase/quadrature}
\acrodef{IM-DD}{intensity-modulated direct-detection}
\acrodef{PD}{photodetector}
\acrodef{OMA}{optical modulation amplitude}
\acrodef{ER}{extinction ratio}
\acrodef{FSR}{free spectral range}
\acrodef{TPA}{two-photon absorption}
\acrodef{FCA}{free carrier absorption}
\acrodef{FCD}{free carrier dispersion}
\acrodef{DWDM}{dense wavelength-division multiplexing}
\acrodef{B2B}{back-to-back}
\acrodef{SP}{single polarization}
\acrodef{DP}{dual polarization}
\acrodef{OSNR}{optical signal-to-noise ratio}
\acrodef{ASE}{amplified spontaneous emission}

\acrodef{OBPF}{optical band-pass filter}

\acrodef{VNA}{vector network analyzer}
\acrodef{ECL}{external cavity laser}
\acrodef{EDFA}{Erbium doped fiber amplifier}

\acrodef{PDME}{polarization division multiplexing emulator}
\acrodef{PC}{polarization controller}
\acrodef{VOA}{variable optical attenuator}
\acrodef{SMF}{single mode fiber}
\acrodef{BPD}{balanced photodetector}
\acrodef{RTO}{real-time oscilloscope}
\acrodef{OSA}{optical spectrum analyzer}
\acrodef{SDM}{space division multiplexing}
\acrodef{WDM}{wavelength division multiplexing}
\acrodef{PAM}{pulse-amplitude modulation}
\acrodef{ASK}{amplitude-shift keying}
\acrodef{BPSK}{binary phase-shift keying}
\acrodef{QAM}{quadrature amplitude modulation}
\acrodef{PRBS}{pseudorandom binary sequence}
\acrodef{SPS}{sample-per-symbol}
\acrodef{ISI}{intersymbol interference}
\acrodef{RF}{radio frequency}
\acrodef{DSP}{digital signal processing}
\acrodef{DPD}{digital pre-distortion}
\acrodef{LUT}{look-up table}
\acrodef{RRC}{root-raised cosine}
\acrodef{FIR}{finite impulse response}
\acrodef{LPF}{low pass filter}
\acrodef{CDE}{chromatic dispersion equalizer}
\acrodef{MIMO}{multiple-input multiple-output}
\acrodef{CMA}{constant modulus algorithm}
\acrodef{MMA}{multi-modulus algorithm}
\acrodef{LMS}{least mean square}
\acrodef{CPR}{carrier phase recovery}
\acrodef{FOC}{frequency offset compensation}
\acrodef{BER}{bit error rate}
\acrodef{FEC}{forward error correction}
\acrodef{EVM}{error vector magnitude}
\acrodef{QPSK}{quadrature phase shift keying}
\acrodef{CPO}{co-packaged optics}
\acrodef{SOI}{silicon-on-insulator}
\acrodef{FOM}{figures of merit}
\acrodef{IM}{intensity modulation}
\acrodef{AM}{amplitude modulation}
\acrodef{PM}{phase modulation}
\acrodef{PCB}{printed circuit board}
\acrodef{OFC}{optical frequency comb}
\acrodef{CW}{continuous wave}
\acrodef{SOA}{semiconductor optical amplifier}
\acrodef{PDFA}{praseodymium-doped fiber amplifier}
\acrodef{BDFA}{bismuth-doped fiber amplifier}
\acrodef{PBS}{polarization beam splitter}
\acrodef{LO}{local oscillator}
\acrodef{TIA}{transimpedance amplifier}
\acrodef{SD}{soft decision}
\acrodef{HD}{hard decision}
\acrodef{CD}{chromatic dispersion}
\acrodef{QD}{quantum dot}
\acrodef{MBE}{molecular beam epitaxy}
\acrodef{DWELL}{dots-in-a-well}
\acrodef{CPM}{colliding-pulse mode-locked}
\acrodef{SA}{saturable absorber}
\acrodef{L-I-V}{light-current-voltage}
\acrodef{IIR}{infinite impulse response}
\acrodef{FIR}{finite impulse response}
\acrodef{RA-MZI}{ring-assisted Mach–Zehnder interferometer}
\acrodef{LPCVD}{Low-Pressure Chemical Vapor Deposition}
\acrodef{OPL}{optical path length}
\acrodef{UID}{unintentionally doped}
\acrodef{SEM}{scanning electron microscope}
\acrodef{HR}{high-reflectivity}
\acrodef{AlN}{aluminum nitride}
\acrodef{SMSR}{side-mode suppression ratio}
\acrodef{PnO}[P\&O]{perturb-and-observe}
\acrodef{FFT}{fast Fourier transform}

\begin{document}

\title[Article Title]{Comb-Driven Coherent Optical Transmitter for Scalable DWDM Interconnects}


\author[1]{\fnm{Alireza} \sur{Geravand}}\email{alireza.geravand.1@ulaval.ca}

\author[1]{\fnm{Erwan} \sur{Weckenmann}}\email{erwan.weckenmann.1@ulaval.ca}

\author[1]{\fnm{Jean-Michel} \sur{Vallée}}\email{jean-michel.vallee.1@ulaval.ca}

\author[1]{\fnm{Farshid} \sur{Shateri}}\email{farshid.shateri.1@ulaval.ca}

\author[1]{\fnm{Zibo} \sur{Zheng}}\email{zibo.zheng.1@ulaval.ca}
\author[1]{\fnm{Simon} \sur{Levasseur}} \email{simon.levasseur@copl.ulaval.ca}
\author[2]{\fnm{Bo} \sur{Yang}}\email{yangbo@iphy.ac.cn}
\author[2]{\fnm{Jiajian} \sur{Chen}}\email{jiajian.chen@iphy.ac.cn}
\author[2]{\fnm{Ting} \sur{Wang}}\email{wangting@iphy.ac.cn}
\author[2]{\fnm{Zihao} \sur{Wang}}\email{wangzihao@iphy.ac.cn}

\author[1]{\fnm{Leslie} \sur{A. Rusch}}\email{leslie.rusch@gel.ulaval.ca}
\author*[1]{\fnm{Wei} \sur{Shi}}\email{wei.shi@gel.ulaval.ca}

\affil[1]{\orgdiv{Department of Electrical and Computer Engineering, Centre d’optique, photonique et laser (COPL)}, 
\orgname{Université~Laval}, \orgaddress{\city{Quebec City}, \state{Quebec}, \country{Canada}}}

\affil[2]{\orgdiv{Beijing National Laboratory for Condensed Matter Physics, Institute of Physics}, 
\orgname{Chinese Academy of Sciences}, \orgaddress{\city{Beijing}, \country{China}}}




\abstract{Driven by the growing demand for large-scale artificial intelligence applications, disaggregated compute nodes and high-radix switches in next-generation computing clusters are set to surpass the capacity of current optical interconnect technologies. Such a surge turns several aspects of transmitters into critical bottlenecks: shoreline bandwidth density and energy efficiency are effectively limiting the scalability. We present a comb-driven coherent optical transmitter architecture on a Si/SiN platform that provides the bandwidth density, energy efficiency, and compact footprint required for such co-packaged-enabled optical interconnects. We evaluate scalability through critical building blocks, including ultra-compact microring-assisted Mach–Zehnder modulators (MRA-MZMs) and dense wavelength-division multiplexing (DWDM) interleavers. Single-tone experiments demonstrate a net line rate of 400~Gbps per polarization (16-QAM, 120~GBd) in silicon within the O-band, achieving a record shoreline density of 4~Tbps/mm while consuming only 10~fJ/bit for modulation. We also demonstrate transmission rates of up to 160~GBd QPSK in back-to-back and 100~GBd over 7~km of fiber without dispersion compensation. Using a quantum-dot frequency comb, six 100~GHz-spaced WDM channels transmit 1.08~Tb/s over 5~km. 
System-level analyses show that by leveraging advanced modulation formats through the integration of wavelength and polarization multiplexing, our proposed architecture can realistically support combined transmission rates exceeding 10~Tbps per fiber within practical limits of power consumption and packaging, outlining a clear path toward future petabit-scale interconnects.
}

\maketitle
\newpage
\section*{Introduction}\label{secIntro}

Data movement between processors has become a critical bottleneck in hyperscale data centers and high-performance computing (HPC) clusters, with network bandwidth trailing computational throughput and off-chip I/O energy rivaling computation itself \cite{Sze_2020,Narayanan2021}. Driven by the rapid growth of AI and machine learning workloads, conventional electrical interconnects can no longer deliver the bandwidth or energy efficiency required at scale \cite{Rajbhandari2021,goldstein2023generative,Sevilla2022}.

Silicon photonics, as a promising platform for \ac{AI} hardware\cite{Tossoun2024,Rizzo2023,Daudlin2025,Yuan2024,Peng2024}
is seen as crucial for the success of \ac{CPO} and the scaling of disaggregated high-performance computing \cite{Shekhar2024}.
While photonic interconnects based on \ac{IM-DD} links are being examined to tackle the challenges of such large-scale, disaggregated systems\cite{Cheng2018,Khani2021}, their limitations restrict the future scalability of such systems.  
The introduction of coherent optical interconnects overcomes the limitations of \ac{IM-DD} based optical interconnects 
\cite{kobayashi2022coherent,Shi2020} by increasing reach and link budget thanks to better receiver sensitivity and robustness against fiber-induced impairments, including chromatic dispersion, polarization mode dispersion, and fiber nonlinear effects \cite{zhou2025dd}. An improved link budget also enables the deployment of optical switching, which enhances the flexibility and efficiency of the clusters.

\begin{figure}[t]%
\centering
\includegraphics[width=0.99\textwidth]{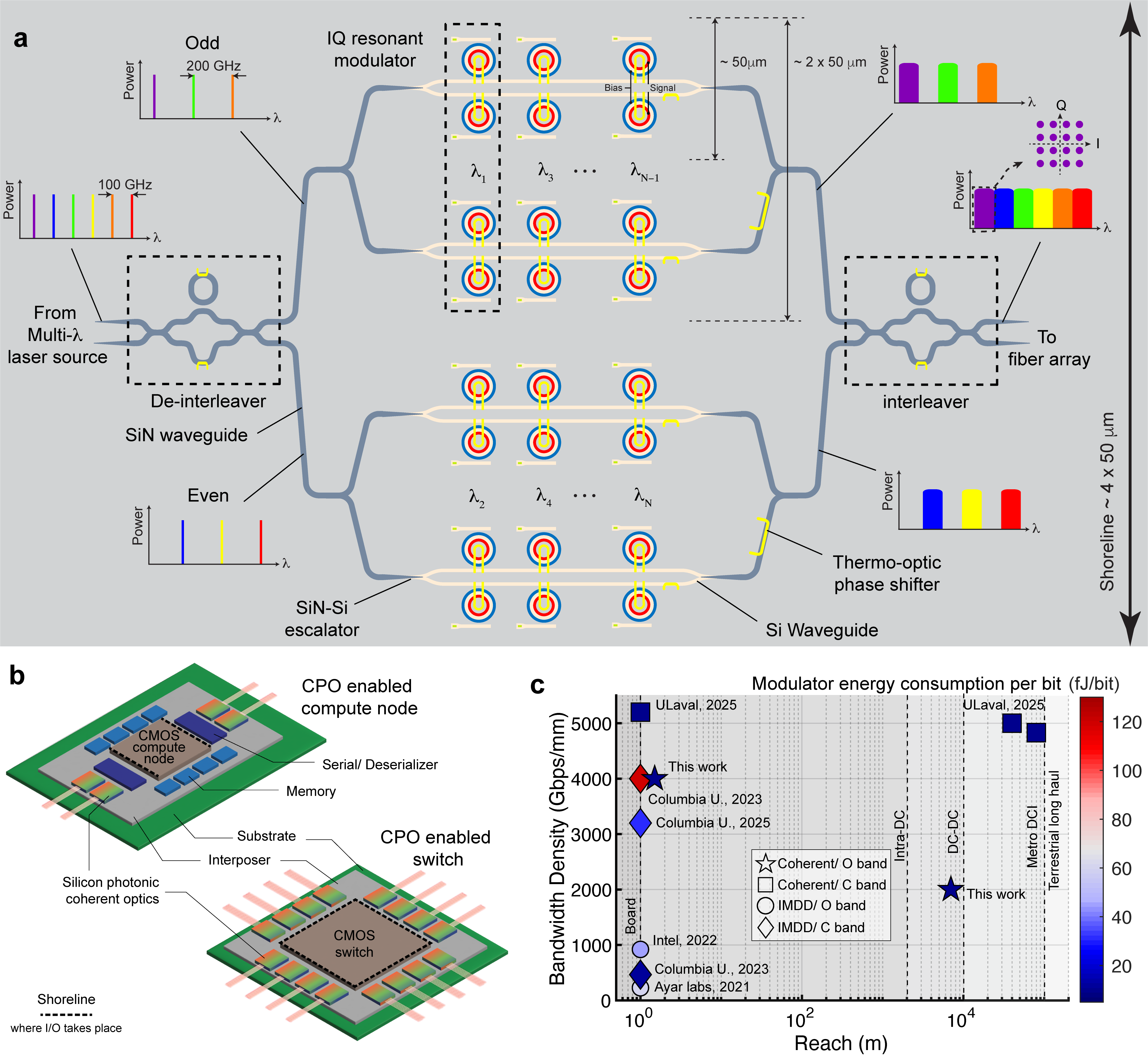}
\caption{\textbf{Conceptualization of scaling comb-driven ultra-compact coherent silicon photonic transmitters.} \textbf{a,} A highly-scalable compact coherent transmitter in the Si/SiN platform exploiting resonator-based, highly-compact modulators with high modulation order and wavelength multiplexing; an optical comb source acting as a multi-\textit{$\lambda$} laser fuels the transmitter. Typical dimensions governing the bandwidth density are indicated.\textbf{b,} Silicon photonic coherent optics at the heart of AI hardware in the form of CPO enabled compute node and switch with CMOS chip shoreline highlighted. 
\textbf{c,} Performance comparison of our device and other state-of-the-art highly compact transmitters in O and C bands \cite{Fathololoumi2022,Sun2020,Daudlin2025,Rizzo2023,Wang2023,Geravand2025} in terms of shoreline bandwidth density, reach, and modulation energy consumption per bit.
}\label{fig:Arch_intro}
\end{figure}

Traditionally designed around millimeter-scale Mach–Zehnder modulators (MZMs), coherent optical transmitters have historically been considered bulky and unsuitable for high-density photonic integration. However, the introduction of \ac{MRM}-based coherent interconnects has addressed these limitations by offering substantial reductions in footprint and power consumption \cite{Geravand2025}. This paved the way toward the tight integration of compact, coherent interconnects and \ac{CMOS} platforms, enabling coherent optical links for switches and compute nodes through \ac{CPO}, as illustrated in Fig.~\ref{fig:Arch_intro}b.

The shoreline (Fig.~\ref{fig:Arch_intro}b) refers to the linear dimension along the interface between electrical and electro-optical devices. \Acp{MRA-MZM}, due to their extremely compact size, take up minimal space and support very high bandwidth densities at chip or package interfaces (Fig.~\ref{fig:Arch_intro}a), commonly referred to as shoreline bandwidth (Tb/s per mm).
The intrinsic wavelength-selective characteristics of \acp{MRM} allow cascading multiple rings into a multi-wavelength transmitter configuration designed for \ac{DWDM}. In such a configuration, each ring independently modulates a distinct wavelength channel.

Silicon-based coherent optical transmitters operating in the C-band have been extensively investigated for long-distance communication. In contrast, there have been relatively fewer demonstrations in the O-band \cite{Misak2025,Bernal2024}, which is more suitable for short-reach applications such as \ac{AI} datacenters and \acp{HPC}. This gap is especially pronounced for compact coherent transmitters in the O-band, which remains underexplored. Recent \ac{MRM}-based transmitters have demonstrated data rates of only up to 200~Gbps per wavelength and polarization using \ac{QPSK} at 120~GBaud \cite{Geravand2025JSTQE}.

Achieving maximum on-chip transmission capacity at scale requires expanding across multiple dimensions, including modulation format, spatial channels, and wavelength channels—all of which can be done with a scalable transmitter architecture. While optimizing individual device performance remains essential, ultimate transmission capacity is governed by system-level integration metrics, scaling primarily with the number of wavelength channels, spatial lanes, and polarization states.
Due to their compact footprint, high electro-optic bandwidth, and intrinsic compatibility with \ac{DWDM} systems, MRA-MZM-based transmitters offer considerable advantages, significantly enhancing scalability across these critical dimensions.

In this work, we introduce a comb-driven scalable coherent optical transmitter architecture implemented on a standard Si/SiN platform, delivering the bandwidth density, energy efficiency, and ultra-compact form factor essential for dense coherent optical interconnects. Through systematic evaluation of critical building blocks, including ultra-compact \ac{I/Q} MRA-MZMs and \ac{DWDM} flat-top interleavers, we demonstrate record-setting performance in the O-band. Single-channel experiments achieve a net line rate of 400 Gb/s per polarization (16-QAM at 120 GBaud) with \ac{BER} below 20\% \ac{FEC} threshold, establishing a record shoreline density of 4~Tb/s/mm while consuming 10~fJ/bit. Transmission of 160~GBd QPSK in back-to-back and 100~GBaud over 7~km fiber without dispersion compensation were also demonstrated at this performance level. Leveraging a quantum-dot frequency comb, we successfully transmit 1.08~Tb/s over 5~km using six DWDM channels spaced at 100~GHz intervals (BER <20\% \ac{FEC}). System-level analyses indicate that, through the integration of wavelength and polarization multiplexing, our proposed architecture can support aggregate transmission rates exceeding 10~Tb/s per fiber within practical constraints of power consumption and packaging. These results, summarized in Fig.~\ref{fig:Arch_intro}c, represent among the highest demonstrated performance metrics in the literature to date \cite{Fathololoumi2022,Sun2020,Daudlin2025,Rizzo2023,Wang2023,Geravand2025}.

\section*{Results}
\subsection*{Scalable architecture}\label{sec:Architecture}
A solution based on resonant devices such as \acp{MRA-MZM} employs a bus architecture.
A single bus architecture faces increased insertion loss as more rings are added, suffers from inter-channel crosstalk, and has a limited number of wavelength channels due to the \ac{FSR}, all of which restrict scalability.
To address the scalability limits of dense coherent transmitters, we propose a multi-bus design incorporating flat-top \ac{SiN} interleavers. 
This interleaved configuration mitigates inter-channel crosstalk and reduces insertion loss from excessive MRM cascading. Unlike silicon, \ac{SiN} exhibits negligible \ac{TPA} and lower propagation loss, substantially improving power handling—key to realizing highly scalable \ac{DWDM} transmitters\cite{Qiu2025,Moss2013,Tan2018,tpa_sin1}. Moreover, cascading such interleavers enables capacity scaling beyond the \ac{FSR} if multi-FSR techniques\cite{James2023,Novick2023} are used.

The proposed scalable, ultra-compact transmitter is fed by a multi-wavelength comb laser, as illustrated in Fig.~\ref{fig:Arch_intro}a. 
The uniformly spaced injected laser tones are separated into even and odd subgroups by the SiN flat-top de-interleaver and routed into two SiN waveguide buses. Each bus integrates compact MRA-MZM modulators fabricated from silicon waveguides, connected to the SiN waveguides via SiN/Si escalators. As the optical power is de-interleaved and split across the buses, the average power in the silicon waveguides remains lower than in the SiN input waveguides, which facilitates high-power operation of the proposed transmitter. The cascaded \ac{I/Q} MRA-MZMs, operating in single-drive push-pull mode, modulate the interleaved tones with data. De-interleaving before modulation increases the spectral spacing between adjacent modulators, effectively eliminating dynamic interchannel crosstalk stemming from the operation of adjacent channels. Additionally, the doubling of the effective inter-channel spacing of MRMs on each bus eliminates the static filtering effect of adjacent MRMs. Moreover, halving the number of modulators on each bus reduces the cumulative insertion loss from the rings, which is particularly beneficial when high doping densities are used to achieve high bandwidth and efficiency. After reconversion to SiN waveguides, the channels are merged by the SiN interleaver, producing tightly spaced, modulated wavelengths that couple into a fiber for transmission to the receiver. 

Although the architecture is symmetric and identical in design, the modulation parameters—such as format and data rate—can be adjusted flexibly and independently for each wavelength channel. While demonstrated with a single level of interleaving for one polarization in Fig.~\ref{fig:Arch_intro}a, this approach can be extended to support two orthogonal polarizations and multiple levels of interleaving. Further details on the interleaver are presented in Methods and Supplementary Note~\ref{SN:sec_Interleaver}.

The shoreline for the proposed multi-bus solution is shown in Fig.~\ref{fig:Arch_intro}a on the far right. Adding wavelength channels increases the device length but does not expand the shoreline; it instead improves the bandwidth density. A pair of push--pull MRMs within an MRA--MZM typically occupies a width of about 
$\sim 50~\mu\text{m}$ (Fig.~\ref{fig:Arch_intro}a), primarily set by the ring diameter. 
An \ac{I/Q} modulator, which integrates two such pairs, requires approximately 
$\sim 100~\mu\text{m}$. Extending this to an interleaved multi-bus transmitter doubles 
the width to $\sim 200~\mu\text{m}$. Notably, this footprint---equivalent to the 
shoreline in our architecture---matches the pitch of standard fiber arrays, enabling 
straightforward scalability by simply adding more fibers in parallel.

\begin{figure}[t]%
\centering
\includegraphics[width=0.97\textwidth]{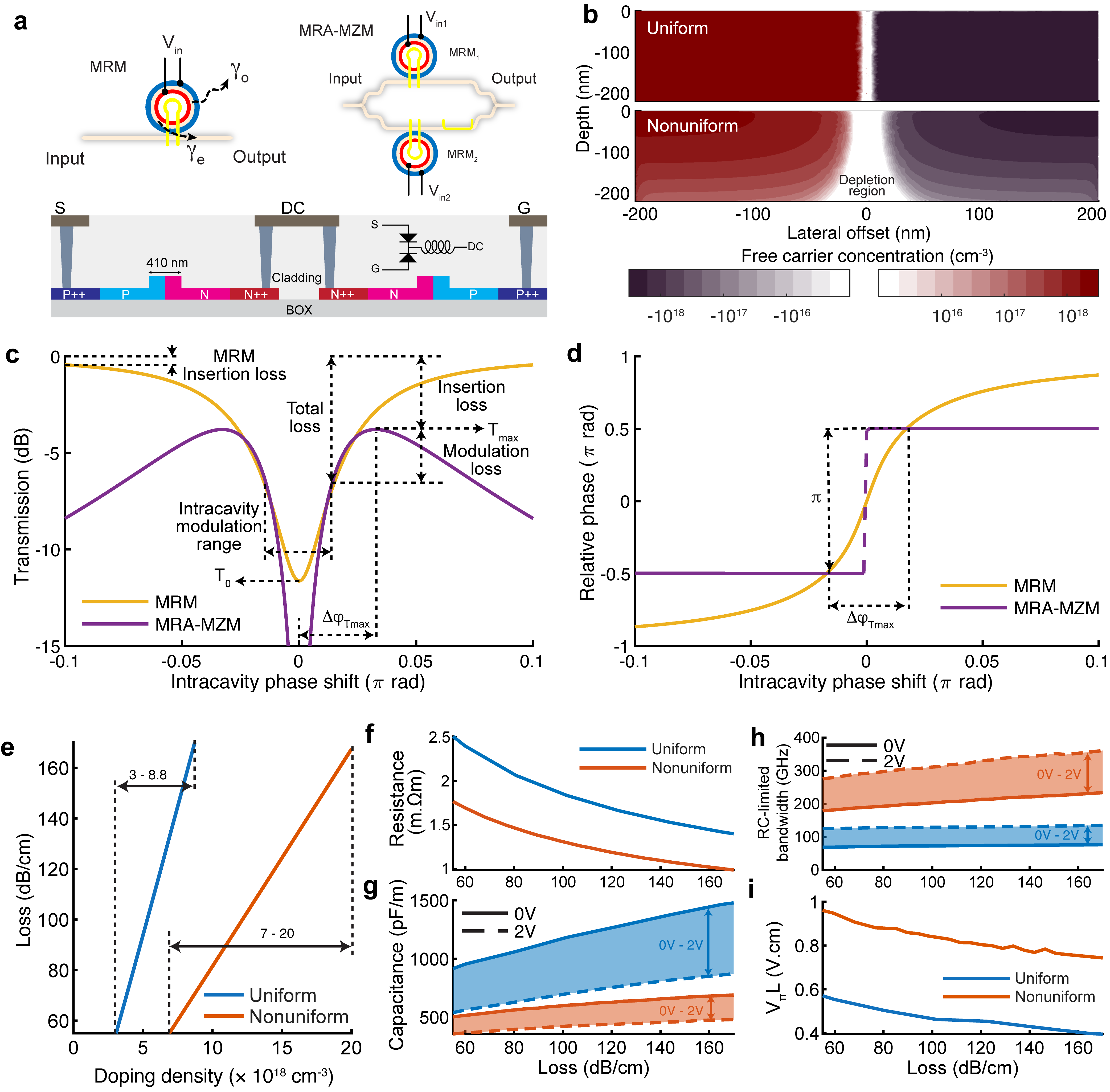}
\caption{\textbf{Principle of MRM and MRA-MZM.} 
\textbf{a,} Schematic diagrams of MRM, MRA-MZM, and the cross section of the PN junction loaded waveguide. \textbf{b,} Free carrier concentration profiles inside the core of the waveguide for uniform and nonuniform doping profiles. \textbf{c,d,} Transmission \textbf{(c)} and relative phase \textbf{(d)} response of MRM and MRA-MZM vs. intracavity phase shift, defining the optical loss components and operation requirements. Maximum transmission of MRA-MZM and the resonance depeth of MRM are indicated as \text{$T_{max}$} and \text{$T_0$}, respectively. \textbf{e,} Optical loss of the waveguide as a function of doping density. \textbf{f,g,h,} Resistance per unit length\textbf{(f)}, capacitance per unit length\textbf{(g)}, and RC-limited bandwidth of the PN junction loaded waveguide\textbf{(h)} for a range of reverse-bias applied voltages as a function of the waveguide loss. \textbf{i,} Half-wave voltage (V$_\pi$L) as a function of waveguide loss at 0.5~V reverse-bias voltage.
}\label{fig:mrm_intro}
\end{figure}

\subsection*{Scalability of MRA-MZM}\label{sec:BWscaling}
Trade-offs between bandwidth, modulation efficiency, and loss in the MRM largely determine the scalability of \ac{I/Q} MRA-MZMs, the critical building blocks of the circuit. These trade-offs are governed by device physics and the material and platform constraints. We first examine the performance metrics of MRMs and MRA-MZMs, focusing on the conventional single-mode, overcoupled add-drop configuration (Fig.~\ref{fig:mrm_intro}a). In this setup, the drop port is weakly coupled and used for resonance stabilization. The intrinsic and leakage decay rates are denoted by $\gamma_o$ and $\gamma_e$, respectively. Overcoupling is assumed, as it enables full complex plane coverage, which is required for coherent optical transmission \cite{Geravand2025}.

In this configuration, the transmission and phase response of the MRM at resonance are modulated by the intracavity phase shift induced by carrier depletion in the PN-junction-loaded silicon waveguide. The PN-junction phase shifter can be implemented in various ways, depending on doping levels and profiles (Fig.\ref{fig:mrm_intro}b). Two representative cases are considered: a uniformly doped abrupt junction and a non-uniform, smoothly graded junction. The latter features a gradual doping concentration across the depletion region with a Gaussian profile peaking near the surface, matching experimental junction profiles. Both designs share a common slab waveguide cross-section with a core thickness of 220~nm and a width of 410~nm. The free-carrier concentration profiles of the core for these two cases are shown in Fig.\ref{fig:mrm_intro}b.

The MRM transmission exhibits a resonance depth (\text{$T_0$}) at resonance and a smooth 
2$\pi$ phase shift across the resonance, characteristic of the overcoupled regime (yellow curves in Fig.\ref{fig:mrm_intro}c–d). This differentially driven intracavity phase shift in resonance-aligned MRMs modulates the transmission of the MRA-MZM, which operates at its null point (with a $\pi$ phase difference between the two arms), as shown by the purple curve in Fig.\ref{fig:mrm_intro}c. Unlike the MRM, the MRA-MZM phase retains only two discrete states ($\pm\frac{\pi}{2}$) due to symmetric differential operation, ensuring chirp-free modulation \cite{Geravand2025}.

The total transmission loss of an MRA-MZM includes two components: optical insertion loss and modulation loss. The optical insertion loss is determined entirely by the device design, while the modulation loss depends on both design parameters, such as \text{$V_{\pi}$} (the voltage required to achieve maximum transmission, (\text{$T_{max}$}) ) and operational factors like the peak-to-peak voltage swing ($V_{\mathrm{pp,diff}} = \max(V_{\mathrm{diff}}) - \min(V_{\mathrm{diff}})$, {$V_\mathrm{diff}(t)=V_\mathrm{in1}(t)-V_\mathrm{in2}(t)$}) that drives the intracavity phase shift (Fig.\ref{fig:mrm_intro}c). To achieve maximum transmission (\textbf{$T_{max}$}), the required intracavity phase shift (\text{$\Delta\varphi_{Tmax}$}) corresponds to each MRM contributing a \text{$\pm\frac{\pi}{2}$} phase shift at the through port with opposite signs (see Fig.\ref{fig:mrm_intro}d). Notably, Fig.~\ref{fig:mrm_intro}c also illustrates that the resonance depth of the MRM (\text{$T_0$}) does not directly determine the insertion loss of the MRA-MZM, contrary to common assumptions.

The unequal overlap of the optical mode with carriers in the two PN-junction profiles results in different levels of optical loss and modulation efficiency. To achieve a target optical loss, an appropriate doping density must be chosen for each profile, as shown in Fig.~\ref{fig:mrm_intro}e.

The doping density of the PN junction affects both the junction capacitance and resistance, while the applied reverse bias significantly modifies the capacitance (see Figs.~\ref{fig:mrm_intro}f–g). As the resistance and capacitance exhibit opposing trends, the RC-limited electrical bandwidth remains nearly constant within the specified range (Fig.~\ref{fig:mrm_intro}h). However, variations in reverse bias voltage alter the capacitance, shift the RC-limited bandwidth, and introduce dynamic nonlinearities.  

The nearly flat trend of electrical bandwidth versus waveguide loss—representing the doping density—indicates that increasing the doping density within such levels does not significantly improve the RC-limited bandwidth. Instead, the junction profile along with the doping density defines the electrical bandwidth.

Furthermore, Fig.~\ref{fig:mrm_intro}i shows the half-wave voltage (\text{$V_\pi L$}), where higher optical loss—caused by increased doping density—correlates with improved modulation efficiency. The nonuniform junction more accurately reflects a practical device and is assumed for the remainder of this work.  

\begin{figure}[]%
\centering
\includegraphics[width=0.99\textwidth]{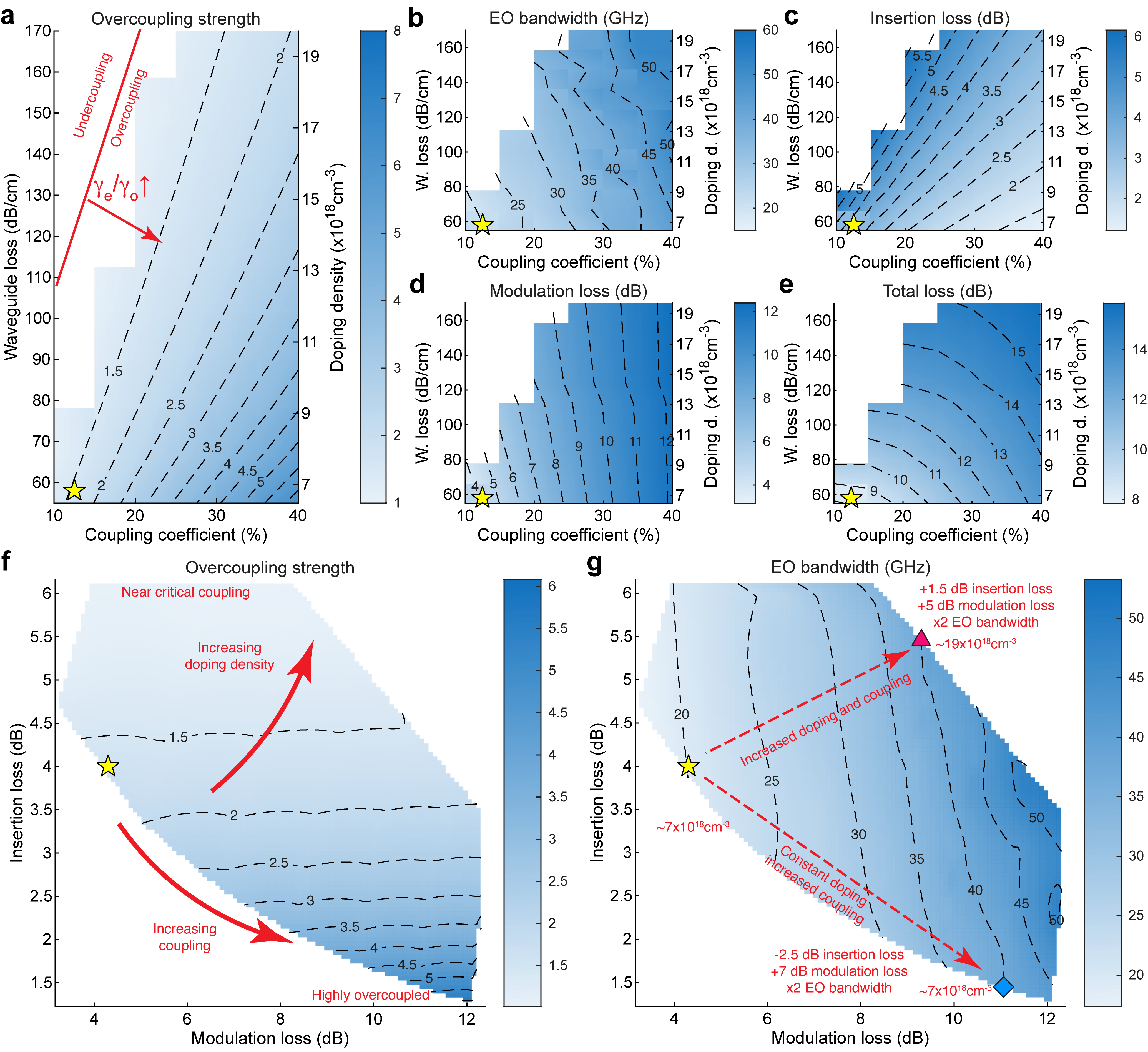}
\caption{\textbf{MRA-MZM scalability.} \textbf{a,} MRM overcoupling strength as a function of waveguide loss/doping density and ring-bus coupling coefficient for a ring with 10~$\mathrm{\mu}$m radius. The doping density is reported for the junction with nonuniform profile. The red line indicates the critical coupling condition, and the omitted regions are under- or critically coupled. \textbf{b,c,d,e,} MRA-MZM electro-optic bandwidth at 2~V bias\textbf{(b)}, insertion loss\textbf{(c)}, modulation loss driven by 2~$\mathrm{V_{pp}}$\textbf{(d)}, and total loss\textbf{(e)} as a function of waveguide loss/doping density and ring-bus coupling coefficient for 10~$\mathrm{\mu}$m radius. \textbf{f,g,} MRA-MZM overcoupling strength\textbf{(f)} and EO bandwidth\textbf{(g)} as a function of the performance metrics: insertion loss and modulation loss when driven by a swing of 2~$\mathrm{V_{pp}}$, showing the trade-offs in design of the modulator. The proof-of-concept device is marked with a yellow star. Devices marked with the blue triangle and the pink square represent other possibilities. The blue square shows a device with the same doping density as the yellow star one, but with stronger coupling. The pink triangle shows a device with higher doping and stronger coupling. Both blue square and pink triangle devices achieve twice the EO bandwidth of the yellow device.
}\label{fig:mrm_sweep}
\end{figure}

High-speed optical modulators are typically evaluated based on their loss and bandwidth. Our study extends to the exploration of MRM and MRA-MZM performance metrics within their design space. Under a range of overcoupled conditions made by independently varying the ring-bus coupling coefficient and loss for rings with 10~$\mu$m radii, the performance metrics are evaluated. 

Assuming an overcoupled condition, the resonance depth can be calculated through \cite{Liang2021}
\begin{equation}
    T_{0} = \left( \frac{\gamma_e/\gamma_o -1}{\gamma_e/\gamma_o +1}\right)^2 = \left( \frac{\gamma_e - \gamma_o}{\gamma_e + \gamma_o}\right)^2,
    \label{eq:res_depth}
\end{equation}
\noindent
where \text{$\gamma_e$} is the extrinsic decay rate of the resonator due to coupling to the bus, \text{$\gamma_o$} is the intrinsic decay rate of the resonator due to the losses, and \text{${\gamma_e}/{\gamma_o}$} indicates the overcoupling strength (e.g. \text{${\gamma_e}/{\gamma_o}>1$} indicates the overcoupling regime). Therefore, the resonance depth of the MRM reveals the depth of the overcoupling condition, which is presented in Fig.~\ref{fig:mrm_sweep}a. Additionally, in the highly overcoupled regime, the optical field decay rate ( \text{$\gamma = \gamma_e + \gamma_o$}, \text{$\tau = 1/\gamma$} )
is primarily determined by the coupling to the bus (i.e. \text{$\gamma \approx \gamma_e$}) 
thus the optical bandwidth is strongly governed by the coupling coefficient. Assuming an operation with zero laser-resonance detuning, the electro-optic bandwidth of MRA-MZM can be simplified to a two-pole system, typically with a dominant optical pole through 
\begin{equation}
    \left( \frac{1}{f_{EO}} \right)^2 = \left( \frac{1}{f_{E}} \right)^2 + \left( \frac{1}{f_{O}} \right)^2 = \left(2\pi RC\right)^2 + \left(2\pi\tau\right)^2, 
    \label{eq:EOBW}
\end{equation}
\noindent
where C, R, and \text{$\tau$} are junction capacitance, resistance, and optical field lifetime \cite{Gheorma2002} of the resonator, as illustrated in Fig.~\ref{fig:mrm_sweep}b. 
The optical insertion loss of the MRA-MZM enhances with increasing overcoupling depth, reflecting the overcoupling trend (presented in Fig.~\ref{fig:mrm_sweep}c). Conversely, the modulation loss has a near inverse relationship with coupling coefficient within the design space, as presented by modulation loss in Fig.~\ref{fig:mrm_sweep}d, assuming a 2~V peak-to-peak swing. The total loss is then determined by the summation of the modulation loss and insertion loss, which is presented in Fig.~\ref{fig:mrm_sweep}e. Notably, the EO bandwidth and total loss display contrasting optimal design parameters within the design space, highlighting a crucial trade-off between these two performance metrics.

Meeting the minimum performance specs in design requires navigating multi-dimensional trade-offs among insertion loss, modulation loss, and EO bandwidth. Rather than plotting each output versus a single independent parameter (coupling coefficient and doping density), we construct surf plots of insertion loss, modulation loss, and EO bandwidth against each other, with coupling strength and doping density as the sweep parameters. These surfaces clarify how improving one metric degrades the others in MRA-MZM design, as shown in Fig.~\ref{fig:mrm_sweep}f-g. A near critically-coupled MRM offers the lowest modulation loss, but it has a high insertion loss.
Additionally, EO bandwidth and modulation loss exhibit opposing trends.
The yellow star in Fig.~\ref{fig:mrm_sweep} marks the proof-of-concept ultra-compact O-band \ac{I/Q} modulator demonstrated in this work, selected to minimize modulation loss while maintaining reasonable insertion loss. The pink triangle and blue square represent alternative designs that trade off insertion and modulation loss to achieve twice the EO bandwidth. Although both devices exhibit similar bandwidths, they differ in doping density: the blue square device shares the same doping as the proof-of-concept, while the pink triangle device has a higher level. As a result, the blue square device operates in a highly overcoupled regime, whereas the pink triangle device is near critical coupling.

\begin{figure}[h]%
\centering
\includegraphics[width=0.9\textwidth]{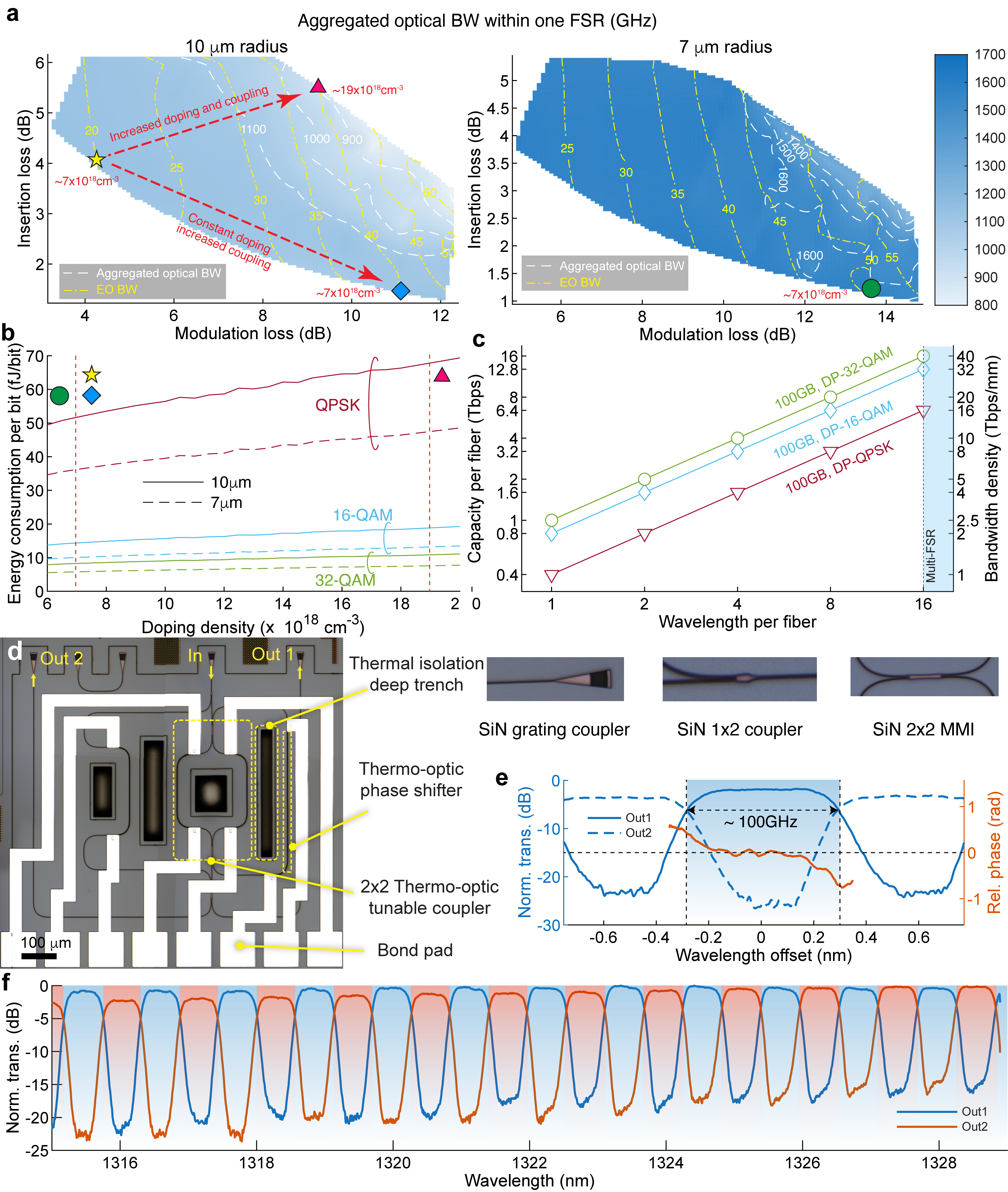}
\caption{\textbf{Scaling ultra-compact transmitters.} \textbf{a,} Aggregated optical bandwidth within one FSR ($\Delta f_\mathrm{agg}$) defined by Eq.~\ref{eq:channel} as a function of modulation loss, insertion loss for 10 and 7~$\mu$m radii. Yellow dashed contours indicate the EO bandwidth of the modulator. Proof-of-concept device and three alternative designs are marked with yellow star, pink triangle, blue square, and green circle, respectively. 
\textbf{b,} Modulation energy consumption per bit as a function of waveguide doping density for various modulation formats and radii, assuming \text{$2~V_{pp}$} swing. \textbf{c,} Capacity scaling of the transmitter per fiber and projected bandwidth density.
\textbf{d,} Microscope images of the fabricated SiN interleaver with annotations and insets of its critical building blocks.\textbf{e,} Measured transmission and relative phase response of the SiN interleaver. \textbf{f,} Wavelength dependency of the measured normalized transmission spectra of the interleaver in a wide span within the O-band. 
}\label{fig:scale}
\end{figure}

\subsection*{Transmitter scalability}\label{sec:transmitter_scaling}

To evaluate the scalability of the MRA-MZM based transmitter architecture, we consider a single-fiber-coupled transmitter (shown in Fig.~\ref{fig:Arch_intro}a). As the bandwidth of the modulator is limited, maximizing transmission capacity and bandwidth density requires multiplexing across wavelengths and orthogonal polarizations. For wavelength multiplexing, the spectral channel count of the transmitter is increased (presented in Fig.~\ref{fig:Arch_intro}a). However, the \ac{FSR} and the off-resonance insertion loss of the MRM ($\text{IL}_{\text{off}}$, see Extended Data Fig.~\ref{fig:extended_fig}a-b) impose fundamental constraints. We define the aggregated optical bandwidth per polarization as the total bandwidth of the maximum number of channels supported within a single \ac{FSR}, assuming sufficient inter-channel isolation and negligible crosstalk. This assumption is justified because the transmitter architecture employs flat-top interleavers, and spectrally adjacent channels are routed on separate buses, thereby minimizing inter-channel crosstalk arising from the static and dynamic filtering of adjacent channels. Assuming Nyquist WDM channels with a zero guard band (i.e. \ac{FWHM} channel spacing) \cite{lin2018frequency}, the wavelength channel count $N_\lambda$ and aggregated optical bandwidth $\Delta f_\mathrm{agg}$ within one \ac{FSR} is estimated by \cite{James2023}:
\begin{equation}
    \Delta f_\mathrm{agg} = \mathrm{FWHM} \times N_\lambda = 
    \mathrm{FWHM} \times
    \left\lfloor \operatorname{min}\left( \underbrace{\frac{\text{FSR}}{\text{FWHM}}}_{\text{Finesse}}, \frac{\text{IL}_{\text{limit}}}{\text{IL}_{\text{off}}}+1 \right)\right\rfloor,
    \label{eq:channel}
\end{equation}
\noindent
where $\text{IL}_{\text{limit}}$ is the maximum loss difference between the first and last MRMs on the bus. The first term in Eq.~\ref{eq:channel} captures the limitation imposed by spectral overlap between adjacent modulated channels after interleaving to a single waveguide.

Fig.~\ref{fig:scale}a shows that the aggregated optical bandwidth per polarization within one \ac{FSR} (black contour lines) defined in Eq.~\ref{eq:channel} exhibits a near-inverse trend to the EO bandwidth (red contour lines), with higher values for rings of lower EO bandwidth due to a lower $\mathrm{IL_{off}}$ (see Extended Data Fig.~\ref{fig:extended_fig}a-b). This effect is particularly pronounced in smaller-radius rings, which offer larger FSRs. Such behavior promotes the maximization of aggregated transmission capacity by employing \ac{DWDM} with cascaded microrings, as presented in Fig.~\ref{fig:Arch_intro}a. The notable difference in aggregated optical bandwidth between the devices marked by the blue square and pink triangle in Fig.~\ref{fig:scale}a indicates that a highly overcoupled modulator design enables greater transmitter capacity via wavelength multiplexing. Furthermore, the smaller-radius ring with identical doping and coupling parameters (green circle in Fig.~\ref{fig:scale}a, 7~µm radius) maximizes the achievable capacity within a single FSR, supporting the 50~GHz bandwidth required for 100~GBaud transmission.

Energy efficiency is critical for \ac{CPO} systems with tight power budgets. Fig.\ref{fig:scale}b shows the modulation energy consumption per bit versus waveguide doping density at a modest drive voltage of $2V_{pp}$. Owing to their low capacitance, the \ac{I/Q} MRA-MZMs achieve energy consumption  well below 100 fJ/bit, with further reductions possible at higher modulation orders and smaller ring radii––making them well-suited for advanced modulation formats such as 16-QAM and 32-QAM. The weak dependence on doping density reflects the nearly constant capacitance of the nonuniform PN junction (Fig.\ref{fig:mrm_intro}g). Devices with identical doping (blue and yellow) exhibit identical energy consumption, which is lower compared to the pink variant.

We evaluate capacity scaling for the device marked by the green circle, assuming a $7~\mu$m ring, dual polarization, and 100~Gbaud channels. As shown in Fig.\ref{fig:scale}c, this transmitter exceeds 10~Tbps per fiber capacity within the FSR-limited regime, enabled by wavelength and polarization multiplexing with advanced modulation formats. Capacities beyond the FSR limit are also feasible using multi-FSR techniques \cite{James2023,Novick2023}. Based on typical shoreline dimensions (Fig.~\ref{fig:Arch_intro}a), we project the shoreline bandwidth density, also shown in Fig.~\ref{fig:scale}c. The projected achievable bandwidth density surpasses the current state-of-the-art, as shown in Fig.~\ref{fig:Arch_intro}c, and lays a path toward future ultra-compact transmitters with 3D integration, enabling such high bandwidths to escape.

As key components of the proposed architecture, the interleaver and \ac{I/Q} MRA-MZM must be co-designed. To demonstrate spectral scalability, we developed a ring-assisted silicon nitride interleaver for O-band \ac{DWDM}, achieving flat-top passbands and steep roll-off, which are required for \ac{DWDM} schemes. Fig.\ref{fig:scale}d shows microscope images of the fabricated device and its subcircuits. Measured transmission and phase responses (Fig.\ref{fig:scale}e) confirm accurate channel separation with low crosstalk and smooth phase profiles—crucial for coherent links. The broadband response across the O-band (Fig.~\ref{fig:scale}f) ensures compatibility with comb-driven configuration and multi-FSR operation.

\subsection*{Experimental demonstration of comb-driven ultra-compact O-band I/Q modulator}\label{sec_exp}

\begin{figure}[h]%
\centering
\includegraphics[width=0.99\textwidth]{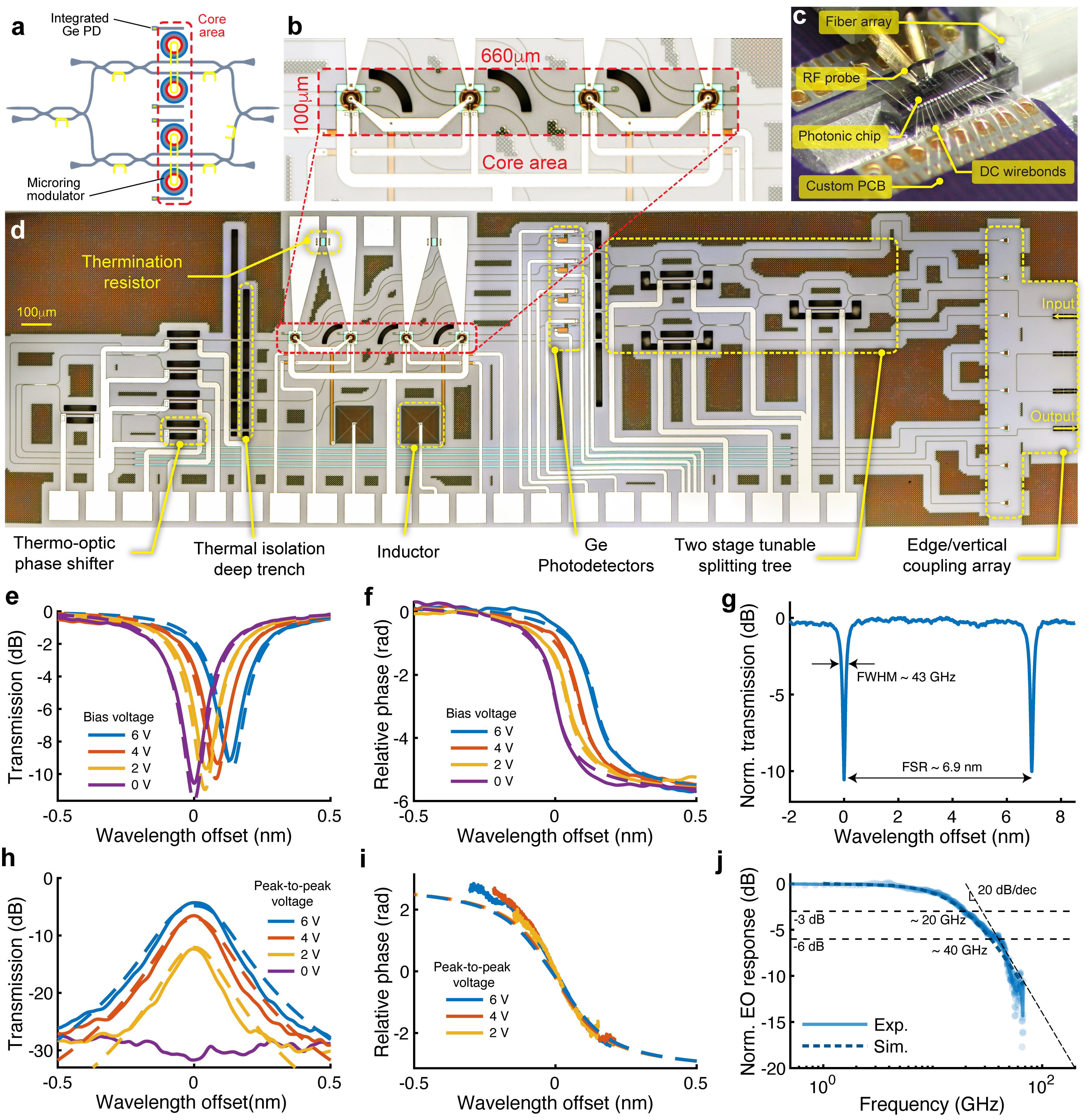}
\caption{\textbf{Ultra-compact O-band IQ modulator and experimental characterization results.} \textbf{a,} Schematic of the demonstrated device. \textbf{b,c,d,} Microscopic images of the device before packaging\textbf{(d)}, the modulator core area\textbf{(b)} and the packaged device under RF probing\textbf{(c)} with annotations indicating the subcomponents and their funcionalities. Red boxes in \textbf{(a,b,d)} indicates the modulation core area with 100~$\mu$m width. \textbf{e,f,} Measured (solid) and simulated (dashed) transmission\textbf{(e)} and phase\textbf{(f)} spectra of the \ac{MRM} with different bias voltages. \textbf{g,} Measured transmission spectra of a solitary MRM at 0~V showing FSR and FWHM. \textbf{h,i,} Measured (solid) and simulated (dashed) normalized transmission\textbf{(h)} and phase\textbf{(i)} spectra of the \ac{MRA-MZM} with different bias voltages. \textbf{j,} Measured normalized small-signal linear electro-optic S\textsubscript{21} responses of the MRA-MZM at zero laser-resonance detuning operating at quadrature point, with dots representing measured data and the solid line representing the smoothed curve.
}\label{fig:exp_res}
\end{figure}

The proposed comb-driven transmitter is experimentally evaluated by designing an ultra-compact O-band \ac{I/Q} modulator, indicated by the yellow star in Fig.\ref{fig:scale}a. The device features a simple \ac{MRA-MZM} architecture based on nested \acp{MZI} (see Methods) and was fabricated using a \ac{CMOS}-compatible silicon process (Fig.\ref{fig:exp_res}a-d). A single-ended, series push-pull drive enables simplified operation with just two \ac{RF} signals and no external bias tees \cite{Geravand2025JSTQE}. The core area footprint (Fig.\ref{fig:exp_res}b) measures \text{$100~\mu m \times 600~\mu m$}, with length primarily determined by RF probe pitch, which can be reduced by optimizing the RF probe configuration. This DWDM-compatible configuration supports multi-wavelength operation by cascading additional ring sets (Fig.\ref{fig:Arch_intro}a), with each comb line modulated by a dedicated \ac{I/Q} modulator.

\subsubsection*{MRM and MRA-MZM characteristics}\label{subsection_DCEO}
The performance metrics of the proposed MRA-MZM were evaluated by characterizing the critical building blocks. This is achieved by experimentally measuring the DC characteristics of individual \acp{MRM} with parameters matching those of the \acp{MRA-MZM}. The transmission and relative phase of the \ac{MRM} at different reverse bias voltages, represented by solid (measured) and dashed (modeled) lines, are shown in Fig.~\ref{fig:exp_res}e,f. The MRM has a resonance depth of about 10.5 dB, a \text{$V_{\pi}$} of $\sim$5~V, and a Q-factor of 5,300. The estimated optical loss in the PN-junction loaded waveguide is $\sim$57 dB/cm. A resonance shift efficiency of 3.4~GHz/V and a modulation efficiency of 1.08~V·cm at a reverse bias of 0.5~V are measured using $V_{\pi}L=\frac{FSR\times\pi R}{\delta\lambda/\delta V}$. The \ac{MRM} shows a \ac{FWHM} of 43~GHz and an FSR of 6.9~nm (or 1.2~THz, see Fig.~\ref{fig:exp_res}g), and the smooth 2$\pi$ phase change across the resonance indicates the overcoupling condition necessary for phase modulation applications. Each MRM has an integrated TiN heater to tune its resonance wavelength with a half-wave power (\text{$P_\pi$}) of 40-50~mW. The thermo-optic shift efficiency of the heaters can be particularly enhanced by thermally isolating the \acp{MRM} from the surrounding area employing deep trenches or by implementing an undercut process \cite{Coenen2022,Parsons2025}.

The DC and \ac{EO} properties of the \ac{MRA-MZM} at the quadrature point are measured and depicted in Fig.~\ref{fig:exp_res}h-j. Figure~\ref{fig:exp_res}i,j illustrate the experimental (solid line) and modeled (dashed line) transmission and phase of the \ac{MRA-MZM} across various driving voltages, showing an extinction ratio of nearly 25~dB at resonance along with chirp-free operation. The normalized small-signal \ac{EO} response of the \ac{MRA-MZM} at the quadrature point is demonstrated in Fig.~\ref{fig:exp_res}j, highlighting bandwidths of 20~GHz and 40~GHz at 3~dB and 6~dB, respectively, which correlate well with the FWHM obtained from the DC measurements of the \ac{MRM}. The roll-off of the \ac{EO} response approximates a 20~dB/dec characteristic of a dual-pole system with a dominant pole. This roll-off enables the device to operate at higher rates, albeit at the cost of requiring stronger pre-compensation.

\subsubsection*{Coherent transmission performance}\label{DataTranssubsection}

\begin{figure}[]%
\centering
\includegraphics[width=0.99\textwidth]{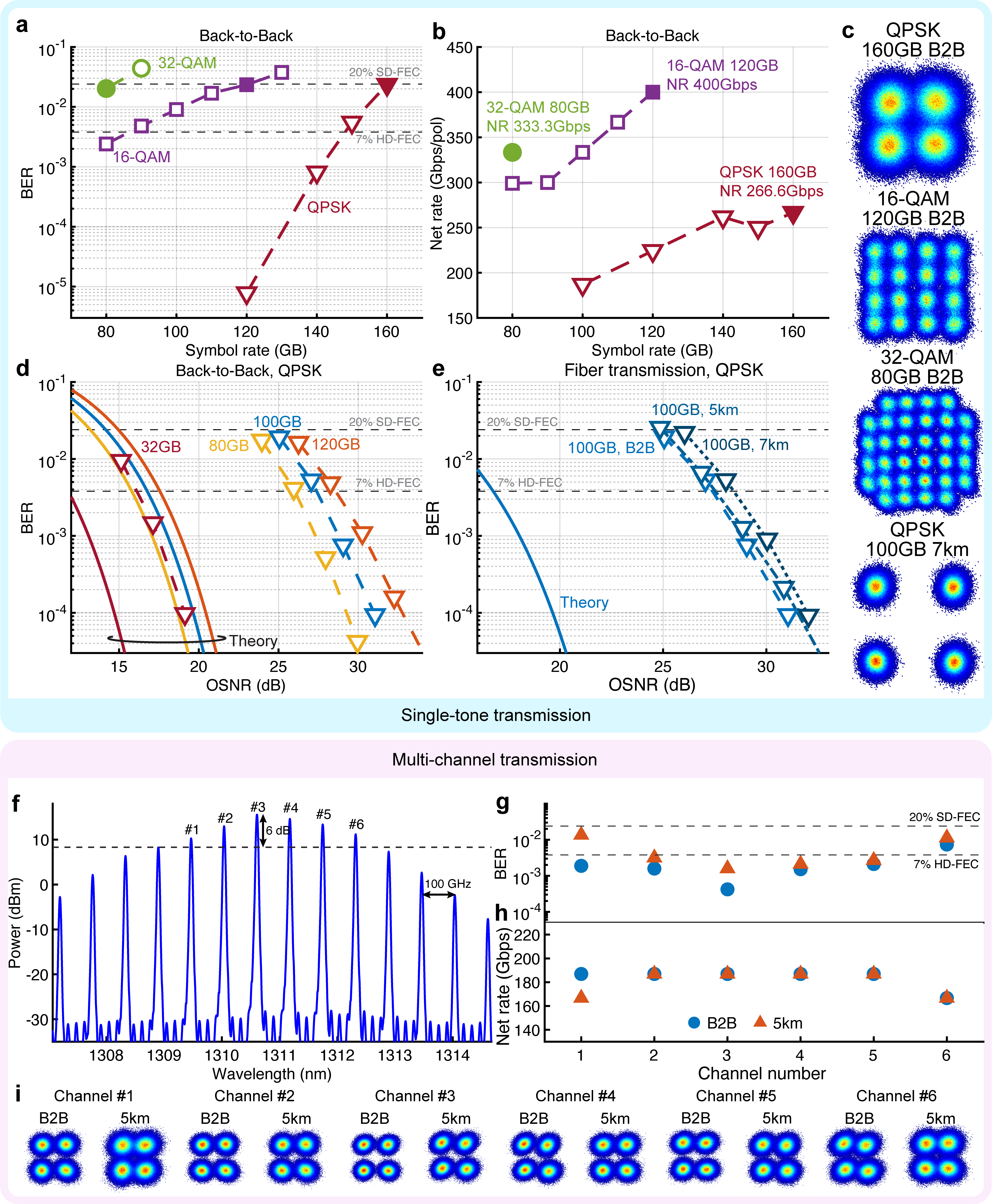}
\caption{\textbf{Coherent data transmission results.} \textbf{a,b,c,d,e,} Data transmission experiment results driven by a single tone laser. BER\textbf{(a)} and net rate\textbf{(b)} vs. symbol rate for QPSK, 16QAM, and 32QAM with typical constellations\textbf{(c)}. Shaded markers indicate the highest achieved net rates for each modulation format. BER vs. OSNR curves for QPSK in back-to-back\textbf{(d)} and fiber transmission\textbf{(e)}. \textbf{f,g,h,i,} Multi-channel coherent transmission results. Amplified comb lines before modulation\textbf{(f)}, measured BER for back-to-back and 5~km configurations\textbf{(g)}, and the resulting net rate for each channel\textbf{(h)} and their reconstructed constellation diagrams\textbf{(i)} are also plotted.
}\label{fig:exp_trans_res}
\end{figure}

Coherent transmission was evaluated in two configurations. In the single-tone setup, a \ac{CW} laser source maximizes power on a single line, enabling assessment of per-channel performance independent of power constraints. In the multi-channel configuration, a quantum dot comb source was used to demonstrate the modulator wavelength selectivity and the comb suitability for \ac{DWDM} interconnects. The comb had six lines, each with lower power than the single-tone setup;  comb-line  power spanned a 6~dB range.  Transmission was performed in both \ac{B2B} and 7~km \ac{SMF} links using a conventional amplified coherent receiver. The setup is detailed in the Methods and shown in Extended Data Fig.\ref{fig:trans_exp_setup}; results are presented in Fig.\ref{fig:exp_trans_res}. Preliminary findings were reported in \cite{Geravand2025_IEEESiP,Geravand2024_ACP}.

In the single-tone, back-to-back configuration, we tested \ac{QPSK}, 16-\ac{QAM}, and 32-\ac{QAM} formats at symbol rates up to 160~Gbaud (Fig.\ref{fig:exp_trans_res}a). At 160~Gbaud, \ac{QPSK} achieves a \ac{BER} below the 20\% overhead \ac{SD}-\ac{FEC} threshold, demonstrating the modulator high-speed operation enabled by its smooth electro-optic roll-off. For 16-\ac{QAM} and 32-\ac{QAM}, lower symbol rates were used to optimize the net bit rate. As shown in Fig.\ref{fig:exp_trans_res}b, the highest net bit rate—400~Gbps—is achieved with 16-\ac{QAM} at 120~Gbaud, marking the highest reported transmission rate per wavelength and polarization for a silicon O-band microring modulator~\cite{Geravand2025_IEEESiP}. This corresponds to a bandwidth density of 4~Tbps/mm, given the \text{$100~\mu m$} device width, comparable to the state-of-the-art C-band results~\cite{Zheng2023ECOC,Geravand2025}. 
Representative constellations are shown in Fig.~\ref{fig:exp_trans_res}c.

Fig.\ref{fig:exp_trans_res}d shows \ac{BER} versus \ac{OSNR} for \ac{QPSK} across symbol rates. Curves for theoretical \ac{QPSK} BER are provided for reference. Increasing the symbol rate introduces a bandwidth-induced power penalty, which reaches 10.5~dB at 100~Gbaud. Nonetheless, a \ac{BER} below the 7\% overhead \ac{HD}-\ac{FEC} threshold is maintained for symbol rates up to 120~Gbaud at OSNRs below 30~dB.

\ac{QPSK} transmission at 100~Gbaud was also demonstrated over 5~km and 7~km of \ac{SMF} without \ac{CD} compensation in the \ac{DSP}. Fig.\ref{fig:exp_trans_res}e shows the measured \ac{BER} versus \ac{OSNR}, revealing only 1~dB penalty from back-to-back to 7~km. This minimal degradation confirms the modulator compatibility with \ac{CD} uncompensated links.

In the multi-channel configuration, a quantum-dot laser generates an \ac{OFC} with six lines within a 6~dB power range, spaced by 100~GHz. After amplification by a \ac{SOA}, the peak power reaches 15.14~dBm, with the spectrum shown in Fig.\ref{fig:exp_trans_res}f. Each comb line is locked to its respective \ac{MRM} using a developed closed loop stabilizer (detailed in Supplementary Note~\ref{SN:sec_Stab}), followed by coherent \ac{QPSK} transmission at 100~Gbaud in both back-to-back and 5~km \ac{SMF} configurations. The measured \ac{BER} and net bit rate for each channel are shown in Fig.\ref{fig:exp_trans_res}g–h, with corresponding constellations in Fig.\ref{fig:exp_trans_res}i. The \ac{BER} primarily reflects the per-line \ac{OSNR}, dictated by comb power. Performance after 5~km transmission remains nearly unchanged, with aggregated net bit rates of 1.10~Tb/s (back-to-back) and 1.08Tb/s (5km), confirming negligible impact from chromatic dispersion.
We note that our prototype chip had a fiber-to-chip loss of 13~dB, well above the state-of-the-art commercial coupling. Our multi-channel results highlight that performance is primarily limited by \ac{OSNR}, governed by insertion loss and available input power.

\section*{Summary and Outlook}\label{secDisc}

We propose a comb-driven, ultra-compact transmitter architecture that enables scalable coherent interconnects to overcome the bandwidth bottlenecks of future AI hardware. Targeting the O-band—well-suited for AI datacenters and \acp{HPC}—we demonstrate record transmission rates from a compact silicon modulator, the highest reported shoreline bandwidth density, and one of the best modulation energy efficiencies to date.

Scalability is validated through co-designed and experimentally demonstrated building blocks: the MRA-MZM and the flat-top SiN interleaver. Multi-channel transmission over kilometer-scale, dispersion-uncompensated links confirms the suitability of the approach for dense wavelength division multiplexing.

These results establish a clear path toward petabit-scale coherent interconnects based on compact, energy-efficient devices and comb-based sources. Looking forward, this architecture lays the foundation for integrated photonic engines capable of meeting the extreme bandwidth and efficiency demands of next-generation AI and high-performance computing systems.

\backmatter

\section*{Methods}

\bmhead{Device design and modeling}
Ansys Lumerical finite difference eigenmode solver and finite difference time domain solver are used for designing the Waveguiding structures and to extract the S-parameters, respectively. Electrical properties, such as capacitance and resistance, of the PN junction are numerically modeled using Lumerical DEVICE. The extracted electrical and optical parameters are then used to create time-domain models in Lumerical INTERCONNECT. Time-domain circuit model parameters are matched to those of the measured ones utilizing the experimental data obtained from DC, RF, and EO characterization.

\bmhead{Silicon photonic chip design and fabrication}
Two nested \acp{MRA-MZM} fed by a tunable coupler form the \ac{I/Q} \ac{MRA-MZM}. The same tunable coupler design is also used in the MRA-MRM. It offers flexibility in test and measurement demonstration and can be replaced by fixed-ratio power splitters in practical cases.
The silicon \ac{MRM} used in the proof-of-concept demonstration features an add-drop configuration with a 10~$\mu$m radius and point couplers with a 170~nm gap for the ring-bus coupler region. Silicon waveguides have a 410~nm width for single-mode operation in O-band, while SiN waveguides used for interleavers are 800~nm-wide and 400~nm-thick.

A standard Si/SiN foundry process on a 220~nm-thick \ac{SOI} wafer at Advanced Micro Foundry (AMF) through MPW service provided by CMC Microsystems is used for the fabrication of the designed devices. Six different doping layers form the lateral PN junction loaded in the waveguides with doping levels of $\sim 6\times 10^{18}cm^{-3}$ in the waveguide core. TiN heaters placed atop the \ac{MRM} and waveguides enable the dynamic control of the resonance of each \ac{MRM} and static tuning to accommodate fabrication nonuniformities across the die. Details of the dynamic tuning of the MRM and stabilization are presented in Supplementary Note~\ref{SN:sec_Stab}.

Each set of \ac{MRA-MZM} is driven in a single drive configuration with 50~$\Omega$ termination resistors placed on the bond pads. The bias voltage is applied through an on-chip inductor to minimize the RF signal leakage to the biasing circuit (showcased in Fig.~\ref{fig:exp_res}d). The core area of the modulator, encompassing the four \acp{MRM} and thermal isolation deep trenches, has a 100~$\mu$m width (red rectangle in Fig~\ref{fig:exp_res}b). Details and consideration of the interleaver design are presented in Supplementary Note~\ref{SN:sec_Interleaver}.

\bmhead{Device packaging and probing}
The silicon die was die-bonded to a custom-designed \ac{PCB} placed atop \ac{TEC} to provide thermal stability throughout the experiments. Only DC pads were wire-bonded in this demonstration, as the RF pads were left exposed for RF probing. 
Suspended edge couplers with a pitch matching the pitch of standard SMF-28 fiber array are used for optical coupling to and from the chip. The packaged chip (presented in Fig.~\ref{fig:exp_res}c) exhibits roughly 6~dB of passive loss due to fiber-to-chip coupling ($\sim$5~dB) and on-chip waveguide routing ($\sim$1~dB). The assembled packaged chip for the comb-driven experiment had a higher coupling loss of approximately 13~dB. RF probing is performed using an unterminated 67~GHz GS-SG probe. An optical vector analyzer (LUNA OVA5013) is used to measure the transmission and phase response of the devices under test.

\bmhead{Quantum Dot laser}
The InAs/GaAs \ac{QD} laser was grown by \ac{MBE}. The active region comprises eight layers of InAs \acp{QD} embedded in a \ac{DWELL} structure, with p-type modulation doping applied to the GaAs capping layers to improve high-temperature performance. The device is designed to operate in the O-band, targeting data center interconnects and on-chip optical I/O applications. Owing to their three-dimensional quantum confinement, \ac{QD} lasers exhibit superior thermal stability and a near-zero linewidth enhancement factor, which significantly reduces sensitivity to optical feedback—an essential feature for on-chip integration without the need for bulky optical isolators\cite{Cui2024}. Moreover, the material system is compatible with monolithic integration on silicon photonic circuits, enabling scalable photonic–electronic co-integration \cite{Wei2023}.
To achieve a frequency comb with 100~GHz channel spacing, we employ a \ac{CPM} cavity design incorporating four electrically driven gain sections and three reverse-biased \ac{SA} sections. The SAs are positioned at 1/4, 1/2, and 3/4 intervals along the Fabry–Pérot cavity, establishing a fourth-order colliding-pulse regime that supports the generation of six comb lines within a 6~dB optical bandwidth. Under optimized operating conditions, the device achieves a peak wall-plug efficiency of 12.5\%.
The gain sections are biased using a Thorlabs CLD1015 laser current controller, while a Keysight E36102A voltage source supplies reverse bias for the SA sections. The device temperature is actively stabilized using a \ac{TEC}. Standard \ac{L-I-V} measurements yield a threshold current of 40~mA and a maximum total output power of 70~mW at an injection current of 300~mA. Optical spectra are acquired using an \ac{OSA} across a range of injection currents (0–300~mA) and SA bias voltages (0–6V), enabling the identification of optimal mode-locking conditions and detailed analysis of comb line spacing, spectral flatness, and bandwidth.
More in-depth details are provided in Supplementary Note~\ref{SN:sec_Laser}.

\bmhead{Transmission setup and digital signal processing}
The experimental setup of the data transmission is shown in Extended Data Fig.~\ref{fig:trans_exp_setup}. Two configurations are shown: comb-driven for a multi-channel transmission and the other using a \ac{CW} laser for the single tone demonstration. Each configuration is highlighted in red and blue, respectively. In both configurations, the source was amplified by an \ac{SOA}, resulting in an average optical power of 14~dBm at the modulator input. Then, an isolator was added to prevent any back reflection from the source before injecting it into the chip, for improved stability. The laser used for the single-tone configuration has a linewidth of 150~kHz, and the quantum-dot laser generates comb lines, from which the one with the highest power has a measured linewidth of 2.8~MHz.

We generate I and Q data from a \ac{PRBS} of $2^{25}-1$ bits, truncated to $2^{18}$ samples to fit the  memory of a 80-GHz bandwidth \ac{DAC} sampling at 256~GSa/s (Keysight M8199B). We apply a raised-cosine pulse shaping filter with a roll-off factor of 0.1 to the data. Another \ac{FIR} filter compensates for the frequency response of the \ac{RF} channels. The RF channels include 65-GHz bandwidth amplifiers, allowing for a peak-to-peak voltage of 4~V, which is applied as 2~V peak-to-peak per MRM in the single-drive series push-pull configuration.

The optical receiver consists of a conventional heterodyne coherent detection system. It employs two-stage amplification for enhanced sensitivity and an \ac{OBPF} to filter out-of-band \ac{ASE} noise. The difference between the multi-channel and single-tone configurations is due to the available equipment at the time of each experiment, i.e. two \acp{PDFA} used for the multi-channel configuration and one \ac{BDFA} and one \ac{PDFA} for the single-tone configuration. We measure spectra with an \ac{OSA} by tapping 2\% of the optical signal power. A \ac{PC} and \ac{PBS} are used to align the signal polarization to the one of a 100-kHz linewidth CW laser used as a \ac{LO}, before sending both to a 90-degree optical hybrid. The outputs of the optical hybrid are connected to two 70-GHz \acp{BPD}. A \ac{VOA} is added to the signal side before the hybrid in order to limit the optical power sent to the \acp{BPD}. In the single-tone configuration, another \ac{VOA} is added at the receiver input to sweep \ac{OSNR}. The \ac{OSNR} is measured with the \ac{OSA}, at a reference resolution of 12.5~GHz. Finally, the two signals are sampled by a \ac{RTO}. In the multi-channel configuration, a 63-GHz bandwidth RTO sampling at 160~GSa/s was used (Keysight DSOZ634A). In the single-tone configuration for transmissions at symbol rates exceeding 120~Gbaud, we used a 110-GHz bandwidth RTO (Keysight UXR1102A) sampling at 256~GSa/s.

We apply \ac{DSP} offline  to recover the full constellation and retrieve the \ac{BER} from the signal acquisitions. First, a tenth-order super-Gaussian \ac{LPF} removes the out-of-band noise. The signal is then resampled to two samples per symbol and a $4\times 4$ \ac{MIMO} filter with 133~taps is applied. \Ac{FFT}-based \ac{FOC} and  blind-search \ac{CPR} are performed~\cite{Selmi2009,Zhou2014}. Finally, a post-MIMO filter with 29~taps is applied to mitigate the distortion between I and Q signals.

The stability of the \acp{MRM} and laser resonance detuning was maintained throughout the experiment using a closed-loop control system comprising on-chip Ge photodetectors, TiN microheaters, off-chip \ac{TIA}, and a controller. See the Supplementary Note~\ref{SN:sec_Stab} for further details.

\bmhead{Modulator energy consumption}
Modulator core power consumption is mainly derived from the dissipation of electric power by junction capacitors ($C_j$) on rising transitions. Assuming that a given QAM signal is equidistant, the consumed energy per bit is given by \cite{Geravand2025}
\begin{equation}
E_b = 4C_jV^2_{pp} \frac{\sum_{i=1}^{\sqrt{N}-1}(\sqrt{N}-i) \left( \frac{i}{\sqrt{N}-1}\right)^2}{N\log_2(N)},
\label{eq:Eb}
\end{equation}
where, $V_{pp}$ is the peak-to-peak voltage swing of the driving signal and $N$ is the QAM order. Note that the consumed power is multiplied by the number of the \acp{MRM}. The junction capacitance used in the numerical study is estimated from simulation results (see Fig.~\ref{fig:mrm_intro}h) under the operating condition(average voltage range) and used to calculate the effective power consumed for modulation given a peak-to-peak voltage swing of 2~V(see Fig.~\ref{fig:scale}b). The experimental demonstration has roughly 18~fF of capacitance. Therefore, the effective energy consumed per bit is 10~fJ/bit, and 5.76~fJ/bit for 16QAM and 32QAM, respectively. Comparison of modulator energy consumption per bit to the state-of-the-art high-bandwidth density coherent and \ac{IM-DD} silicon photonic transmitters is depicted in Fig.~\ref{fig:Arch_intro}c, where we only consider energy efficiency of modulators. As system implementations can vary widely, comparison is confined to modulators, which are the focus of this study.

\section*{Data availability}
The data that supports the findings of this study are available from the corresponding authors upon reasonable request.

\section*{Code availability}
The code used in this study is available from the corresponding authors upon reasonable request.

\section*{Acknowledgments}
This work is funded by NSERC (CRDPJ538381-18). 
We thank Nathalie Bacon, Arman Safarnejadian, and David Turgeon for their technical support and CMC Microsystems for access to MPW services.

\section*{Author contributions}
A.G. and W. S. proposed the concept and conceived the chip design. A.G. carried out the simulations, designed the mask layout, and developed the stabilization mechanism. F.S. designed the SiN interleaver and its mask layout. A.G. conducted the optical and DC characterization, assisted by E.W. and F.S. S.L. designed the PCB, contributed to the stabilization mechanism design, and carried out the electrical packaging. A.G. performed electro-optic characterization with assistance from E.W. J.V. characterized the quantum-dot comb laser. E.W., Z.Z., and L.A.R. designed the data transmission experiment, which were led by E.W. with assistance from A.G. 
B.Y., J.C., T.W., and Z.W. designed and fabricated the quantum-dot comb laser.
All authors contributed to the data analysis. A.G. prepared the manuscript with input from all authors.
W.S. and L.A.R. supervised the project.

\section*{Competing interests}
The authors declare no competing interests.

\section*{Additional information}
Additional modeling and experimental results are provided in the Supplementary Notes.









\begin{appendices}






\bigskip
\newpage
\bigskip

\bigskip
\newpage
\bigskip

\begin{Extended Data Fig.}[h]%
\centering
\includegraphics[width=0.99\textwidth]{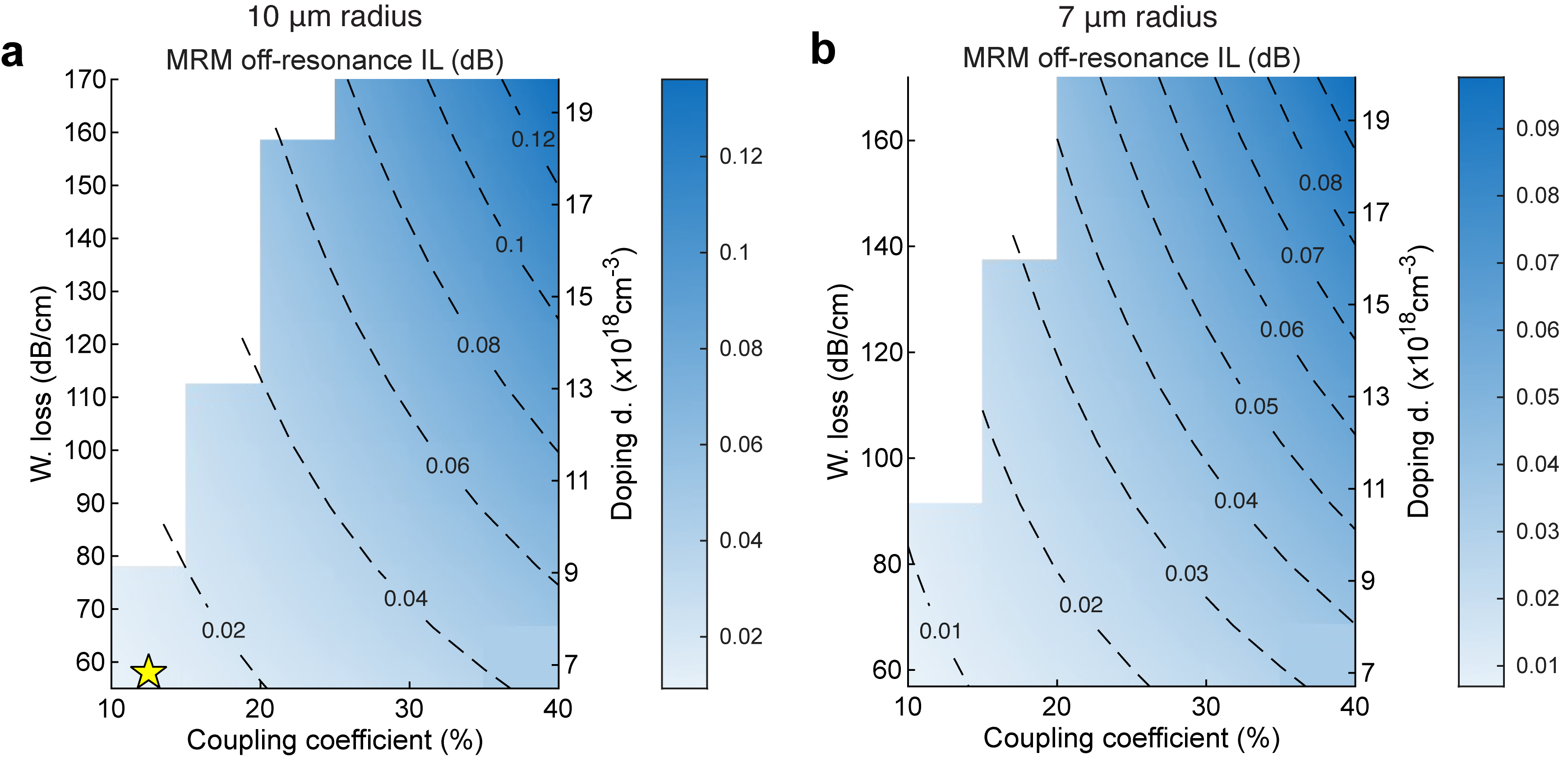}
\caption{\textbf{Design space exploration.} \textbf{a,b,} MRM off-resonance insertion loss within a range of ring-bus coupling strength and waveguide doping densities for $10~\mu m$\textbf{(a)} and $7~\mu m$\textbf{(b)} rings. The star-marked point indicates the proof-of-concept device demonstrated in this work. 
}\label{fig:extended_fig}
\end{Extended Data Fig.}

\bigskip
\newpage
\bigskip

\begin{Extended Data Fig.}[h]%
\centering
\includegraphics[width=0.99\textwidth]{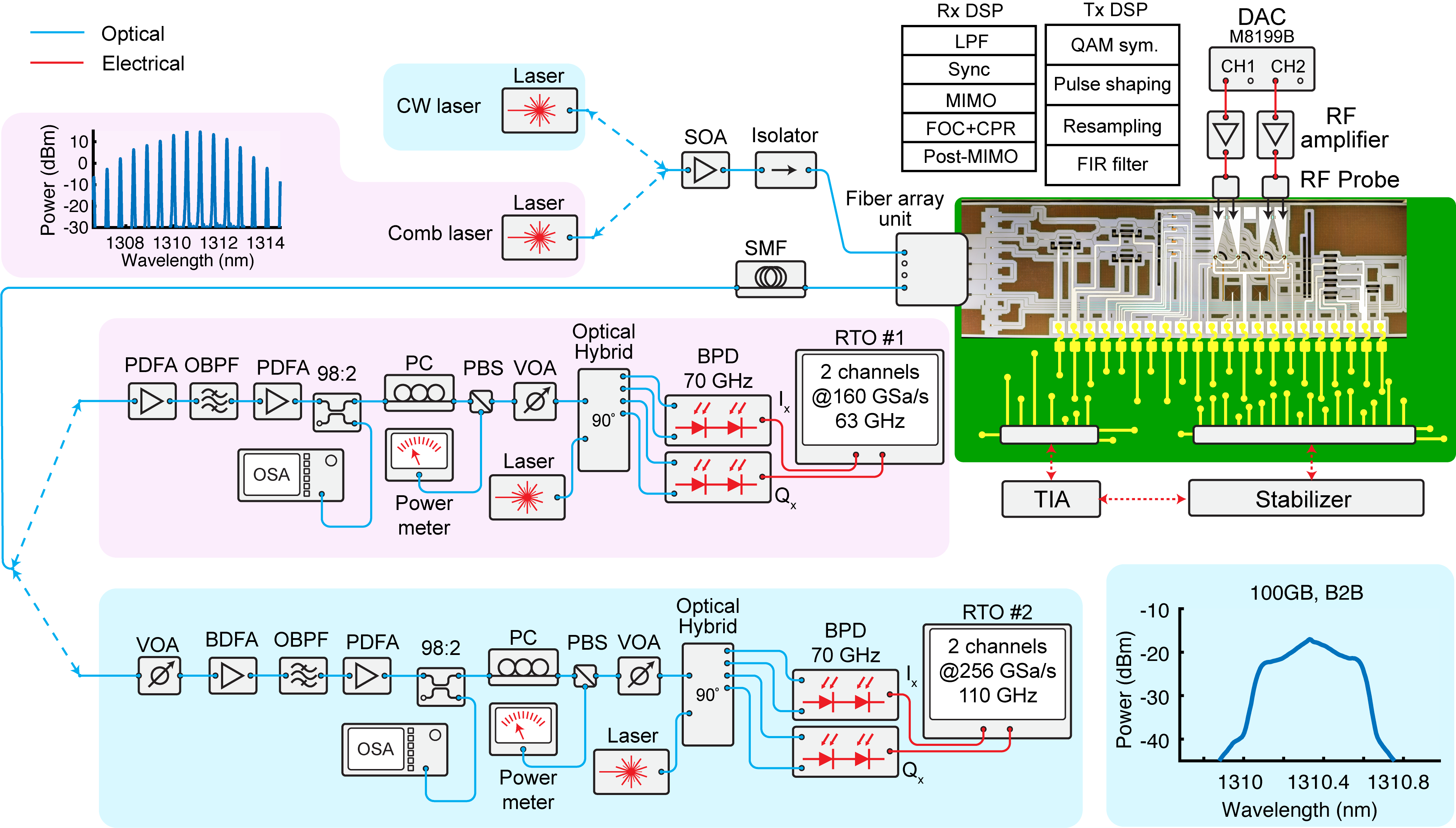}
\caption{\textbf{Data transmission experiment setup.} Experimental setup including the multi-channel transmission configuration, highlighted in pink, and single-tone transmission configuration, highlighted in blue. The silicon chip is represented as packaged on its custom PCB board. Each configuration has a corresponding coherent detection receiver setup. Tx and Rx DSP stacks at the transmitter and receiver are represented. The optical spectrum of the single-tone transmission at 100~Gbaud in QPSK and back-to-back is shown. SOA: semiconductor optical amplifier, DAC: digital-to-analog converter, RF: radio frequency, TIA: trans-impedance amplifier, SMF: single-mode fiber, PDFA: praseodymium-doped fiber amplifier, BDFA: bismuth-doped fiber amplifier, OBPF: optical band-pass filter, OSA: optical spectrum analyzer, PC: polarization controller, PBS: polarization beam splitter, VOA: variable optical attenuator, BPD: balanced photodetector, RTO: real-time oscilloscope, Tx: transmitter, Rx: receiver.
}\label{fig:trans_exp_setup}
\end{Extended Data Fig.}

\newpage
\section*{Supplementary Note}

\pagenumbering{arabic} 
\setcounter{page}{1} 

\renewcommand{\thefigure}{Supplementary Fig. \arabic{figure}}
\setcounter{figure}{0} 
\renewcommand{\figurename}{} 

\startcontents[Supplementary Note]
\printcontents[Supplementary Note]{}{1}{}
\bigskip
\newpage
\bigskip

\section{Supplementary Note: Considerations for the design of the SiN interleaver}\label{SN:sec_Interleaver}

We provide a brief overview of the \ac{SiN} interleavers used in the proposed scalable coherent transmitter. Each interleaver has one input and two spectrally complementary outputs, with the passband of one output aligning with the stopband of the other. As shown in Fig.~\ref{fig:Arch_intro}b, a de-interleaver at the transmitter input separates even and odd wavelengths into distinct modulation buses. This band interleaving serves two purposes: it increases wavelength spacing to reduce inter-channel crosstalk and lowers the number of modulators per bus, thereby minimizing insertion loss \citeS{Rizzo2021}. After modulation, an identical interleaver recombines the channels into a single output. Both interleaver and de-interleaver share the same physical design.

Interleavers can be implemented using either \ac{IIR} or \ac{FIR} architectures. Our design employs a \ac{RA-MZI} interleaver—an \ac{IIR} filter due to the presence of the ring resonator (Fig.\ref{fig:Arch_intro}b, Fig.\ref{fig:scale}d). While \ac{FIR} designs using cascaded MZIs can achieve flat-top responses, they require significantly larger footprints, particularly in \ac{SiN}, where lower index contrast demands larger bend radii. As shown in Fig.~\ref{fig:scale}e, the \ac{RA-MZI} structure achieves a flat-top passband, enabled by the ring’s phase response.

A potential limitation of \ac{RA-MZI} filters in silicon is reduced power handling due to nonlinear effects such as \ac{TPA}, \ac{FCA}, and \ac{FCD}. In contrast, SiN exhibits a negligible \ac{TPA} coefficient owing to its wide bandgap, making it well-suited for high-power \ac{DWDM} applications\citeS{tpa_sin1}.

For a mathematically rigorous analysis of this \ac{RA-MZI} interleaver, one can refer to Chapter 6 of the book by C. Madsen \citeS{madsen_filter}. Here, for the sake of brevity, we only point out the necessary conditions for designing an interleaver. For a RA-MZI structure, in order to realize a flat-top spectral response, the following relations must hold.

\begin{equation}
\begin{aligned}
L_{\text {r }} & =\frac{2 {\lambda_0}^2}{n_g \mathrm{FSR}}, \\
\Delta\mathrm{d} & =\frac{L_{\text {r}}}{2}, \\
\kappa_{c}^2 & = 8/9,
\end{aligned}
\label{eq:interleaver}
\end{equation}
where $\lambda_0$ is the design wavelength, $n_g$ is the group index of the waveguide, $L_{\mathrm{r}}$ is the ring's length, and $\Delta\mathrm{d}$ is the length of the imbalanced arm of the MZI, and $\kappa_c^2$ is the coupling power from MZI to the ring. Note that the value of $\kappa_c^2 = 8/9 \approx 0.89$ could be derived analytically \citeS{madsen_filter}.  
The proposed interleaver in our architecture is based on the SiN platform. There are two main reasons why SiN was preferred for interleaver implementations \citeS{sin_interleaver}. First, SiN waveguides exhibit a low propagation loss of approximately 1 dB/cm, compared to 3 dB/cm for silicon waveguides. This reduced loss is particularly advantageous in multi-stage interleaver configurations, as it minimizes cumulative insertion loss, enabling more efficient and scalable architectures. Second, SiN has negligible two-photon absorption, which significantly enhances the device's power-handling capacity.

\begin{figure}[t]
    \centering
    \includegraphics[width=0.7\linewidth]{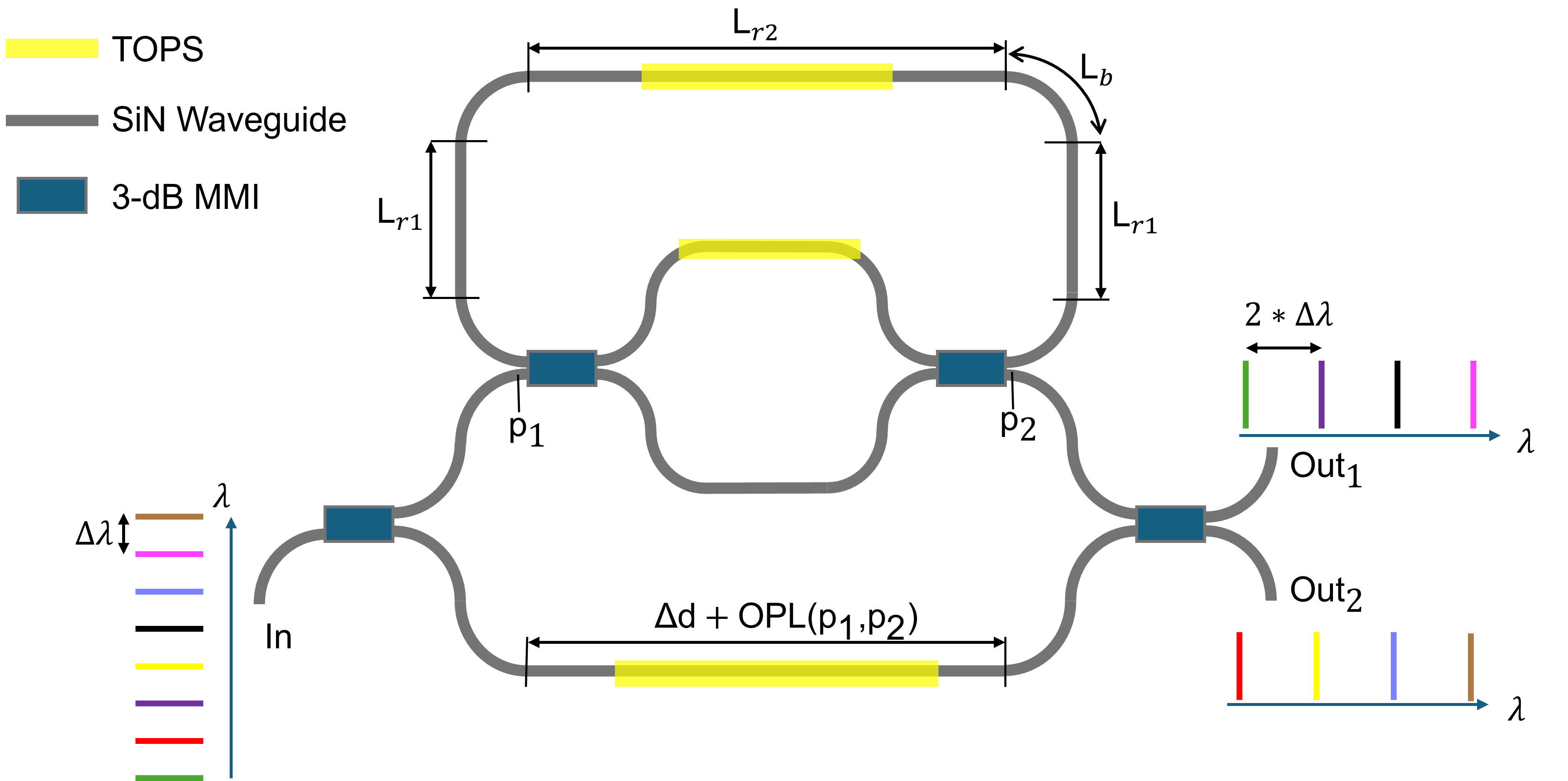}
    \caption{Schematics of the interleaver. This ring assisted filter increases the wavelength spacing by a factor of two. The OPL of the tunable coupler (from $p_1$ to $p_2$) must be subtracted and added to the ring's length ($L_r)$ and $\Delta\mathrm{d}$, respectively.}
    \label{fig:SN_inter_sup}
\end{figure}

The SiN waveguide used for this interleaver is based on a \ac{LPCVD} process with a 400~nm waveguide thickness. To ensure single TE mode, and at the same time, realize relatively sharp waveguide bends, the waveguide width was chosen to be 800~nm. For these waveguide dimensions, the group index of the fundamental TE mode is calculated as $n_g = 2.04$. Given the input laser wavelength spacing, by setting $\mathrm{FSR = 1.144}$~nm the ring's length becomes $L_r = 1470.673$ $\mu{m}$ and $\Delta{d} = L_r/2$. However, it should be noted that these calculations are based on the assumption of having an ideal point coupler with the coupling coefficient indicated by $\kappa_c^2$. Since we have used a tunable coupler realized by a coupler-MZI-coupler structure (Fig.~\ref{fig:scale}d), the \ac{OPL} of the tunable coupler must be subtracted and added to the length of the ring ($L_r$) and the MZI differential section ($\Delta_d$), respectively. This OPL starts from $p_1$ (shown in \ref{fig:SN_inter_sup}) to point $p_2$. The OPL of the tunable coupler, consisting of two 3-dB MMIs and a balanced MZI, is calculated as the sum of the optical path length of the balanced MZI arms and that of the two MMIs at the input and output sections of the tunable coupler. To calculate the path length introduced by MMI, we use the global phase of MMI derived using the self-imaging theory of MMIs \citeS{mmi_globalphase}. The total OPL of the tunable coupler is approximately equal to the OPL of a routing SiN waveguide with dimensions 800~nm $\times$ 400~nm having a length of $487.56$ $\mu m$. We denote this value as $OPL(p_1, p_2)$. In summary, according to the notation in \ref{fig:SN_inter_sup}, the lengths $L_b$, $L_{r1}$, and $L_{r2}$ should be chosen such that $L_r - OPL(p_1, p_2 )= 4L_b + 2L_{r1} + L_{r2} $. 

Note that a $\pi$-phase shift must be added to the ring to align with the phase response of the MZI's differential length. This additional phase shift is provided by a \ac{TOPS}. In addition to this phase shifter, we used two other \ac{TOPS}: one on the tunable MZI-based coupler and the other on the lower arm of the main MZI. The TOPS on the tunable coupler is used to ensure a $\kappa_c^2 = 8/9$ power coupling to the ring, and the one on the MZI is used for post-fabrication tuning. Due to fabrication errors, we might use this TOPS to compensate for the unwanted phase shift generated by fabrication variations such as width and thickness variations of the waveguide. One way to reduce the impact of these variations is to use a wider waveguide width \citeS{widewidth_inter} or multi-cross section designs \citeS{multi_cross_fabtolerant}.

\section{Supplementary Note: Quantum dot comb laser}\label{SN:sec_Laser}
\begin{figure}[th]%
\centering
\includegraphics[width=0.99\textwidth]{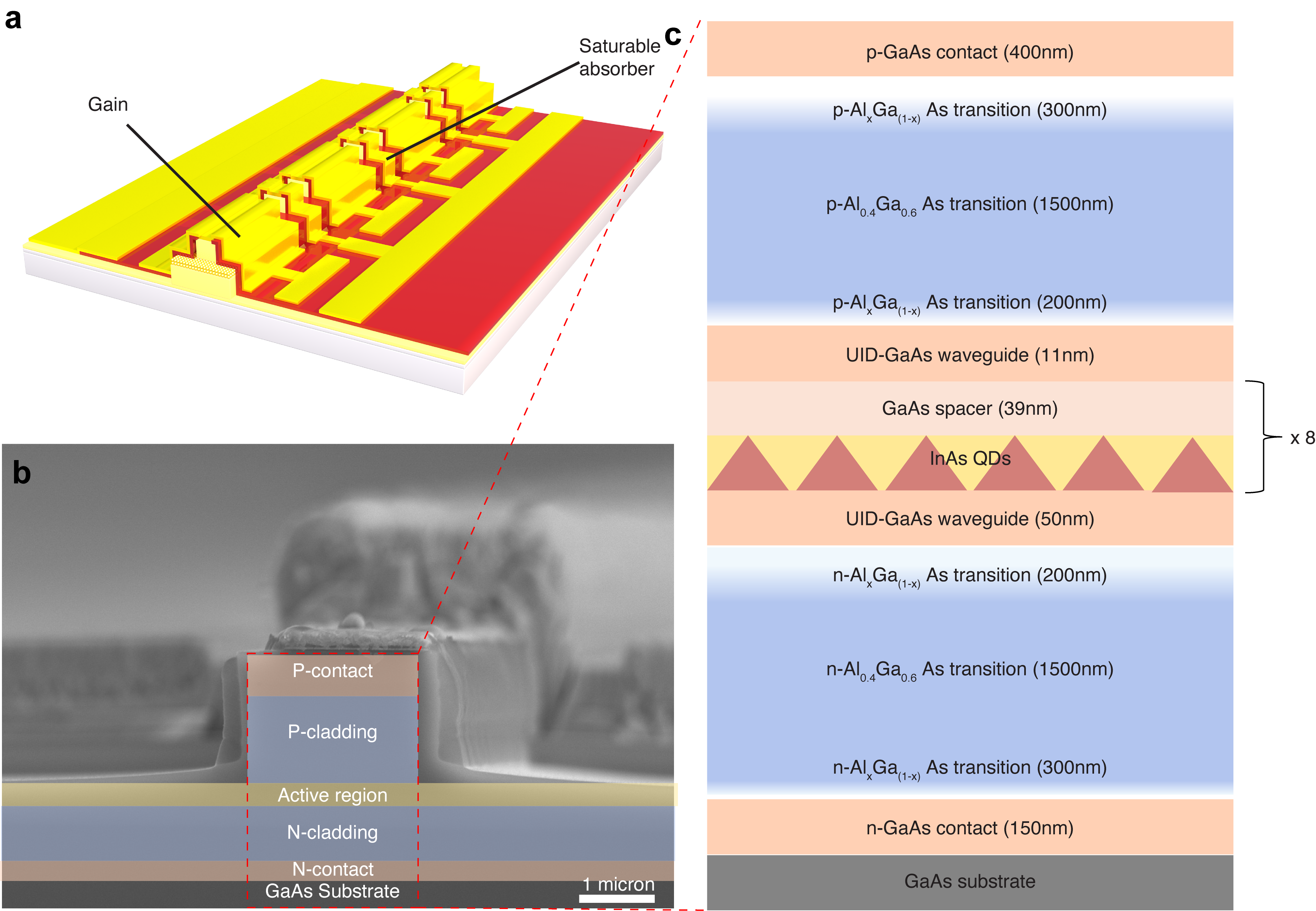}
\caption{\textbf{(a),} Modeling of 4th order 100 GHz QD comb laser. \textbf{(b),} SEM image of QD comb laser. \textbf{(c),} Epitaxial layout of InAs QD laser on GaAs substrate
}\label{fig:SN_laser}
\end{figure}

\subsection{Design Considerations}
A wide channel spacing of 100~GHz is selected for the comb laser source used in this work. Achieving such spacing with conventional two-section mode-locked lasers would require impractically short cavities (e.g., 0.395mm), which are insufficient to provide the optical gain needed for lasing. To overcome the trade-off between 100~GHz comb spacing and output power, a \ac{CPM} scheme is employed, as illustrated in \ref{fig:SN_laser}a. In this configuration, \ac{SA} sections are placed at fractional positions (1/N) along the cavity to support Nth-order harmonic mode-locking, effectively multiplying the cavity length by N and enabling significantly higher output power. In this work, a fourth-order (N=4) CPM structure is implemented, with SAs located at 1/4, 1/2, and 3/4 of the Fabry–Pérot cavity. The cavity is designed with a fundamental round-trip frequency of 25~GHz, such that the fourth harmonic generates a stable frequency comb with 100~GHz channel spacing.

\subsection{Fabrication}
The epitaxial structure of the InAs/GaAs quantum dot (QD) laser was grown using III–V \ac{MBE} and subsequently processed using standard ridge-waveguide semiconductor laser fabrication techniques. A cross-sectional \ac{SEM} image of the laser facet is shown in \ref{fig:SN_laser}b, highlighting the metal contacts, ridge waveguide, and the active region comprising InAs \ac{QD}s embedded between AlGaAs cladding layers.

The full epitaxial layer structure, illustrated in \ref{fig:SN_laser}c, begins with a 150~nm n-type GaAs contact layer, followed by a 300~nm n-type AlGaAs transition layer, a $1.5\mu m$ n-\text{$\mathrm{Al_{0.4}Ga_{0.6}As}$} cladding layer, and a 200nm n-AlGaAs transition layer. A 50nm \ac{UID} GaAs waveguide layer is then deposited to enhance optical confinement. The active region consists of eight layers of self-assembled InAs QDs, each separated by 39~nm GaAs spacer layers. Above the active region, a symmetrical epitaxial sequence completes the structure, comprising a \ac{UID} GaAs layer, C-doped p-type $\mathrm{Al_{0.4}Ga_{0.6}As}$ cladding/transition layers, and a 400~nm p-GaAs contact layer.

Following epitaxy, the standard ridge semiconductor laser fabrication process is then carried out. Ridge waveguides are defined using standard photolithography and dry etching techniques. The mesa and ridge widths are 25$\mu$m and 3$\mu$m, respectively. The total cavity length is $\sim1648\mu$m, corresponding to a fundamental round-trip frequency of 25~GHz. The \ac{SA} sections are each 50$\mu$m long to ensure sufficient absorption for stable mode-locking. Electrical isolation between gain and SA sections is achieved via a narrow dry-etched gap ($\sim$ 10 $\mu$m wide), etched to a depth around 100nm above the active region. This configuration results in an inter-section resistance exceeding 1~k$\Omega$, effectively preventing leakage current. A 500~nm layer of silicon dioxide is deposited across the device surface for passivation. Separate metal contacts are used for the gain and \ac{SA} sections to allow independent electrical biasing. After fabrication, the laser bar is cleaved and one facet is coated with a \ac{HR} coating to enhance output power and reduce threshold current. The laser chip is mounted on an \ac{AlN} carrier to improve thermal management.

\subsection{Characterization Results}
\ac{L-I-V} measurements at 23\textdegree{C} show a threshold current of 40~mA with no roll-over up to 300~mA, achieving up to 70~mW of total output power. Optical spectral mapping identifies optimal operation at a 195 mA injection current, 3 V SA bias, and an operating temperature of 43.5~°C, yielding six comb lines within a 6 dB bandwidth, each exhibiting $>40$~dB \ac{SMSR} and a maximum comb line power of 3.7~dBm. This operating point was chosen to balance maximum power, stability, and spectral purity of the comb lines. These results demonstrate the QD laser's suitability for high-channel-count, low-crosstalk on-chip \ac{DWDM} applications, leveraging the advantages of quantum dots, including a low threshold current, high-temperature stability, and a reduced linewidth enhancement factor compared to quantum well counterparts.

\bigskip
\newpage
\bigskip

\section{Supplementary Note: Stabilization of MRMs}\label{SN:sec_Stab}

To operate at resonance, the MRMs must be stabilized and locked to the laser wavelength. This is achieved via a closed-loop stabilization unit implementing a \ac{PnO} hill-climbing algorithm. The system monitors the photocurrent from drop-port photodetectors after conversion to voltage by a logarithmic \ac{TIA} to track the resonance condition and adjusts the TiN heater atop each MRM to maintain alignment.  

The control algorithm (Algorithm~\ref{alg:control}) is implemented digitally and interfaced to the device under test through \ac{ADC} and \ac{DAC} units. The algorithm consists of two stages: an initial voltage sweep to coarsely align the resonance, followed by \ac{PnO} tracking. The initial sweep finds the heater voltage maximizing the power at the drop port, indicating that the laser-resonance detuning is minimized. The following \ac{PnO} stage tracks and maximizes the power on the drop port. Thermal drift across the chip, caused by other heating elements, shifts the optimal heater power during operation, necessitating continuous tracking. Each MRM is stabilized in parallel.

\begin{figure}[th]%
\centering
\includegraphics[width=0.99\textwidth]{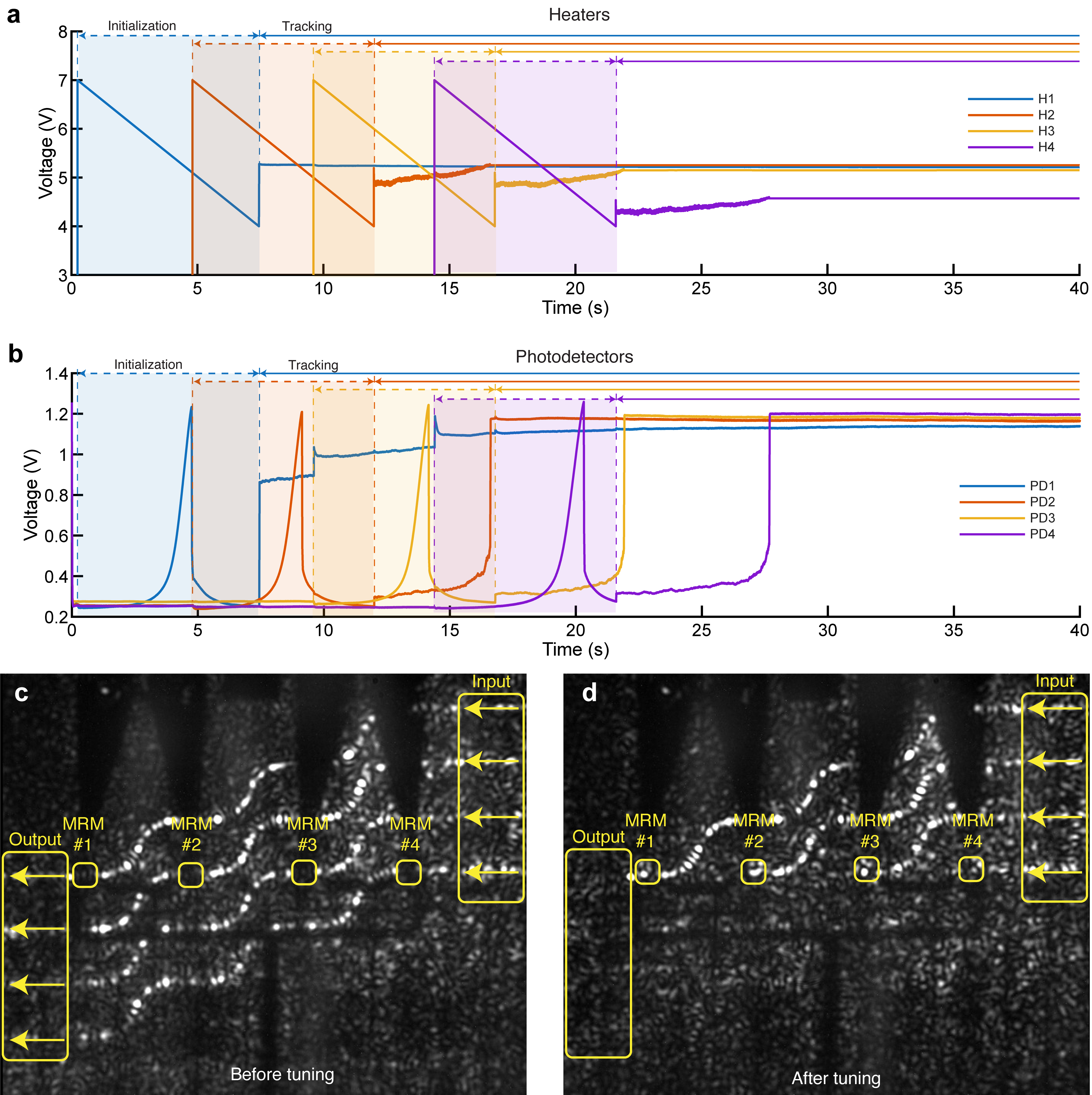}
\caption{\textbf{(a),} recorded heater voltage of MRMs throughout experiment. \textbf{(b),} recorded photodetector voltage of MRMs after TIA throughout the experiment. Micrograph of the laser propagation through waveguides before \textbf{(c)} and after \textbf{(d)} resonance tuning and stabilization.
}\label{fig:SN_stabilizer}
\end{figure}

\ref{fig:SN_stabilizer}a–b shows the heater and photodetector voltages for an \ac{I/Q} MRA-MZM consisting of four MRMs. Rings initialize sequentially, entering the tracking stage immediately after, with shaded regions indicating the initialization phase. Stable and near equal final photodetector voltages confirm proper power balancing in the MZI arms. The tracking algorithm smoothly bypasses occasional local maxima caused by abrupt changes in adjacent heaters. The asymmetric Lorentzian shape of the response at the drop ports reveals the appearance of optical nonlinear behaviours within MRMs.   

\ref{fig:SN_stabilizer}c–d presents top-view micrographs of the demonstrated device before and after tuning, captured by a near-infrared camera. The input laser enters after splitting into four equal branches from the top right. Laser paths from the input to the output appear bright before resonance alignment. After tuning and stabilization, most of the power couples into the rings, darkening the waveguide traces toward the output due to the resonance depth of the rings. 

Other \acp{TOPS} are statically set through the same \ac{DAC} unit and manually tuned throughout the experiment to optimize operation at the null point. Additional details on the stabilization scheme are provided in \citeS{Geravand2025}.

\begin{algorithm}[b]
\caption{Two-Stage Perturb and Observe Control for Microring Modulator}
\begin{algorithmic}[1]
\State \textbf{Stage 1: Initialization by Voltage Sweep}
\State Define voltage sweep range: $V \in [V_{\min}, V_{\max}]$ with step $\delta$
\State Initialize $P_{\text{res}} \gets 0$, $V_\text{res} \gets V_{\max}$
\For{$V$ from $V_{\max}$ to $V_{\min}$ with step $-\delta$}
    \State Apply tuning voltage $V$
    \State Measure output signal $P$
    \If{$P > P_{\text{res}}$}
        \State $P_{\text{res}} \gets P$
        \State $V_\text{res} \gets V$
    \EndIf
\EndFor

\State \textbf{Stage 2: Perturb and Observe Tracking}
\State Set direction $d \gets +1$
\State Set step size $\Delta V$
\State $P_{\text{prev}} \gets P_{\text{res}}$
\While{system is running}
    \State Apply tuning voltage $V \gets V_\text{res} + d \cdot \Delta V$
    \State Measure current output $P$
    \If{$P > P_{\text{prev}}$}
        \State $P_{\text{prev}} \gets P$
        \State $P_{\text{res}} \gets P$
        \State $V_\text{res} \gets V$
        \Comment{Continue in same direction}
    \Else
        \State $d \gets -d$
        \Comment{Reverse direction}
    \EndIf
\EndWhile
\end{algorithmic}
\label{alg:control}
\end{algorithm}

\bibliographystyleS{plain}

\end{appendices}
\end{document}